{
\documentclass[7pt,a4paper]{article}
\pagestyle{empty}
\def\leaderdot{\leaders\hbox to 1 em {\hss.\hss}\hfill}

\dimen0= \parindent                           
\dimen1= \hsize \advance\dimen1 by -\dimen0   

\dimen2=\baselineskip
\def\skiplines#1 { \dimen3=\dimen2 \multiply\dimen3 by #1 \vskip \dimen3}
\def\fullline{\hbox to \fullhsize}

\def\numpage{\baselineskip=24pt\fullline{\the\footline}}

\def\mathcedilla{\vtop{\hbox{c}{\kern0pt\nointerlineskip}
	         {\hbox{$\mkern-2mu \mathchar"0018\mkern-2mu$}}}}

\mathchardef\gq="0060
\mathchardef\dq="0027
\mathchardef\ssmath="19
\mathchardef\aemath="1A
\mathchardef\oemath="1B
\mathchardef\omath="1C
\mathchardef\AEmath="1D
\mathchardef\OEmath="1E
\mathchardef\Omath="1F
\mathchardef\imath="10 
\mathchardef\fmath="0166
\mathchardef\gmath="0167
\mathchardef\vmath="0176

\def\colleft{\strut\kern.3em}
\def\colright{\kern0pt}

\def\figureh{\hbox to}

\def\m@th{\mathsurround=0pt}
\newif\ifdtpt
\def\displ@y{\openup1\jot\m@th
    \everycr{\noalign{\ifdtpt\dt@pfalse
    \vskip-\lineskiplimit \vskip\normallineskiplimit
    \else \penalty\interdisplaylinepenalty \fi}}}
\def\eqalignl#1{\,\vcenter{\openup1\jot\m@th
                \ialign{\strut$\displaystyle{##}$\hfil&
                              $\displaystyle{{}##}$\hfil&
                              $\displaystyle{{}##}$\hfil&
                              $\displaystyle{{}##}$\hfil&
                              $\displaystyle{{}##}$\hfil\crcr#1\crcr}}\,}
\def\eqalignnol#1{\displ@y\tabskip\centering \halign to \displaywidth{
                  $\displaystyle{##}$\hfil\tabskip=0pt &
                  $\displaystyle{{}##}$\hfil\tabskip=0pt &
                  $\displaystyle{{}##}$\hfil\tabskip=0pt &
                  $\displaystyle{{}##}$\hfil\tabskip=0pt &
                  $\displaystyle{{}##}$\hfil\tabskip\centering &
                  \llap{$##$}\tabskip=0pt \crcr#1\crcr}}
\def\leqalignnol#1{\displ@y\tabskip\centering \halign to \displaywidth{
                   $\displaystyle{##}$\hfil\tabskip=0pt &
                   $\displaystyle{{}##}$\hfil\tabskip=0pt &
                   $\displaystyle{{}##}$\hfil\tabskip=0pt &
                   $\displaystyle{{}##}$\hfil\tabskip=0pt &
                   $\displaystyle{{}##}$\hfil\tabskip\centering &
                   \kern-\displaywidth\rlap{$##$}\tabskip=\displaywidth
                   \crcr#1\crcr}}
\def\eqalignc#1{\,\vcenter{\openup1\jot\m@th
                \ialign{\strut\hfil$\displaystyle{##}$\hfil&
                              \hfil$\displaystyle{{}##}$\hfil&
                              \hfil$\displaystyle{{}##}$\hfil&
                              \hfil$\displaystyle{{}##}$\hfil&
                              \hfil$\displaystyle{{}##}$\hfil\crcr#1\crcr}}\,}
\def\eqalignnoc#1{\displ@y\tabskip\centering \halign to \displaywidth{
                  \hfil$\displaystyle{##}$\hfil\tabskip=0pt &
                  \hfil$\displaystyle{{}##}$\hfil\tabskip=0pt &
                  \hfil$\displaystyle{{}##}$\hfil\tabskip=0pt &
                  \hfil$\displaystyle{{}##}$\hfil\tabskip=0pt &
                  \hfil$\displaystyle{{}##}$\hfil\tabskip\centering &
                  \llap{$##$}\tabskip=0pt \crcr#1\crcr}}
\def\leqalignnoc#1{\displ@y\tabskip\centering \halign to \displaywidth{
                  \hfil$\displaystyle{##}$\hfil\tabskip=0pt &
                  \hfil$\displaystyle{{}##}$\hfil\tabskip=0pt &
                  \hfil$\displaystyle{{}##}$\hfil\tabskip=0pt &
                  \hfil$\displaystyle{{}##}$\hfil\tabskip=0pt &
                  \hfil$\displaystyle{{}##}$\hfil\tabskip\centering &
                  \kern-\displaywidth\rlap{$##$}\tabskip=\displaywidth
                  \crcr#1\crcr}}

\def\doublelow#1{\,\vtop{\ialign{\hfil$##$\hfil\crcr
                 \mathstrut #1 \crcr}}\,}
\def\charlvmidlw#1#2{\,\vtop{\ialign{##\crcr
      #1\crcr\noalign{\kern1pt\nointerlineskip}
      $\hfil#2\hfil$\crcr}}\,}
\def\charlvlowlw#1#2{\,\vtop{\ialign{##\crcr
      $\hfil#1\hfil$\crcr\noalign{\kern1pt\nointerlineskip}
      #2\crcr}}\,}
\def\charlvmidup#1#2{\,\vbox{\ialign{##\crcr
      $\hfil#1\hfil$\crcr\noalign{\kern1pt\nointerlineskip}
      #2\crcr}}\,}
\def\charlvupup#1#2{\,\vbox{\ialign{##\crcr
      #1\crcr\noalign{\kern1pt\nointerlineskip}
      $\hfil#2\hfil$\crcr}}\,}

\def\vspce{\kern4pt} \def\hspce{\kern4pt}    

\def\emptybox{\vbox{\kern.7ex\hbox{\kern.5em}\kern.7ex}}
 \font\sevmi  = cmmi7              
    \skewchar\sevmi ='177          
 \font\fivmi  = cmmi5              
    \skewchar\fivmi ='177          
\font\tenmib=cmmib10
\newfam\bfmitfam
\def\bfmit{\fam\bfmitfam\tenmib}
\textfont\bfmitfam=\tenmib
\scriptfont\bfmitfam=\sevmi
\scriptscriptfont\bfmitfam=\fivmi

\def\deltabf{{\bfmit \mathchar"710E}}
\def\epsilonbf{{\bfmit \mathchar"710F}}

\def\xibf{{\bfmit \mathchar"7118}}

\def\sigmabf{{\bfmit \mathchar"711B}}
\def\taubf{{\bfmit \mathchar"711C}}

\def\varepsilonbf{{\bfmit \mathchar"7122}}

\def\twodot{.\kern-0.1em.}

\def\paral{\mathrel{/\kern-.25em/}}
\def\grlo{\mathrel{\hbox{\lower.2ex\hbox{\rlap{$>$}\raise1ex\hbox{$<$}}}}}
\def\logr{\mathrel{\hbox{\lower.2ex\hbox{\rlap{$<$}\raise1ex\hbox{$>$}}}}}
\def\greq{\mathrel{\hbox{\lower1ex\hbox{\rlap{$=$}\raise1.2ex\hbox{$>$}}}}}
\def\loeq{\mathrel{\hbox{\lower1ex\hbox{\rlap{$=$}\raise1.2ex\hbox{$<$}}}}}
\def\grsim{\mathrel{\hbox{\lower1ex\hbox{\rlap{$\sim$}\raise1ex\hbox{$>$}}}}}
\def\losim{\mathrel{\hbox{\lower1ex\hbox{\rlap{$\sim$}\raise1ex\hbox{$<$}}}}}
\font\ninerm=cmr9
\def\uniset{\rlap{\ninerm 1}\kern.15em 1}

\def\emptysq{\mathbin{\vbox{\hrule\hbox{\vrule height1ex \kern.5em 
                            \vrule height1ex}\hrule}}}
\def\emptyrect{\mathbin{\vbox{\hrule\hbox{\vrule height1ex \kern1em 
                              \vrule height1ex}\hrule}}}
\def\rightonleftarrow{\mathrel{\hbox{\raise.5ex\hbox{$\rightarrow$}\ignorespaces
                                   \lower.5ex\hbox{\llap{$\leftarrow$}}}}}
\def\leftonrightarrow{\mathrel{\hbox{\raise.5ex\hbox{$\leftarrow$}\ignorespaces
                                   \lower.5ex\hbox{\llap{$\rightarrow$}}}}}

\def\bkB{{\rm I\kern-.17em B}}
\def\bkC{{\rm \kern.24em
            \vrule width.05em height1.4ex depth-.05ex
            \kern-.26em C}}
\def\bkD{{\rm I\kern-.17em D}}
\def\bkE{{\rm I\kern-.17em E}}
\def\bkF{{\rm I\kern-.17em F}}
\def\bkG{{\rm \kern.24em
            \vrule width.05em height1.4ex depth-.05ex
            \kern-.26em G}}
\def\bkH{{\rm I\kern-.22em H}}
\def\bkI{{\rm I\kern-.22em I}}
\def\bkJ{{\rm \kern.19em
            \vrule width.02em height1.5ex depth0ex
            \kern-.20em J}}
\def\bkK{{\rm I\kern-.22em K}}
\def\bkL{{\rm I\kern-.17em L}}
\def\bkM{{\rm I\kern-.22em M}}
\def\bkN{{\rm I\kern-.20em N}}
\def\bkO{{\rm \kern.24em
            \vrule width.05em height1.4ex depth-.05ex
            \kern-.26em O}}
\def\bkP{{\rm I\kern-.17em P}}
\def\bkQ{{\rm \kern.24em
            \vrule width.05em height1.4ex depth-.05ex
            \kern-.26em Q}}
\def\bkR{{\rm I\kern-.17em R}}
\def\bkT{{\rm \kern.24em
            \vrule width.02em height1.5ex depth 0ex
            \kern-.27em T}}
\def\bkU{{\rm \kern.30em
            \vrule width.02em height1.47ex depth-.05ex
            \kern-.32em U}}
\def\bkZ{{\rm Z\kern-.32em Z}}

\gdef\l{\lambda}

\gdef\m{\mu}

\gdef\s{\sigma}
\gdef\bfs{{\sigmabf}}
\gdef\S{\Sigma}
\gdef\bfS{{\bf \Sigma}}

\gdef\o{\over}

\gdef\bfE{{\bf {E}}}

\gdef\bfx{{\bf {x}}}

\usepackage{epsfig}
\usepackage{oldlfont}
\usepackage{graphics}
\usepackage{oldlfont}

\usepackage{xcolor}

\usepackage{geometry}
\geometry{verbose,
  a4paper,
  tmargin=1.cm,
  bmargin=2.5cm,
  lmargin=2.5cm,
  rmargin=2.cm,
  headheight=0cm,
  headsep=0.cm,
  footskip=0cm}

\def\doublelow#1{\,\vtop{\ialign{\hfil$##$\hfil\crcr
                 \mathstrut #1 \crcr}}\,}
\newcommand{\beg}{\begin{equation}}
\newcommand{\en}{\end{equation}}
\newcommand{\monadress}{ \begin{center} L.M.A./ C.N.R.S. \\
31 Chemin Joseph Aiguier \\
13402. Marseille. Cedex 20. France. \\
\end{center}}
\begin{document}
\title{\bf A numerical method for computing the overall response of nonlinear 
composites with complex microstructure}
\author {H. Moulinec, P. Suquet}
\date{}
\maketitle
\monadress
\begin{abstract}
{\it 
The local and overall responses of nonlinear composites are classically
investigated by the Finite Element Method. We propose an
alternate method based on Fourier series which avoids meshing and which makes
direct use of microstructure images. It is based on the exact expression
of the Green function
of a linear elastic and homogeneous comparison material. 
First the
case of elastic nonhomogeneous constituents is considered and an iterative
procedure is proposed to solve the Lippman-Schwinger equation 
which naturally arises in the problem.  Then, the method is extended to nonlinear
constituents by a step-by-step integration in time. The accuracy of the
method is assessed by varying the spatial resolution of the microstructures. 
The flexibility of the
method allows it to serve for a large variety of microstructures.
}

\end{abstract}

\font\eightrm=cmr8
\font\eightbf=cmbx8
\font\tenrm=cmr10
\font\tenbf=cmbx10
\font\twelverm=cmr12
\font\twelvebf=cmbx12
\font\dixseptrm=cmr17
\hyphenation{iso-tro-pe sui-vante Fou-rier com-po-site permet-tant}
\hyphenation{mo-du-le}
\parindent=8pt

\def\init{\tabskip 0pt\offinterlineskip}
\def\crr{\cr\noalign{\hrule}}


\section{Introduction}

This study is devoted to a numerical method introduced by  
{  Moulinec} and {  Suquet} \cite{MOU94}, \cite{MOU95} to 
determine the local 
and overall responses of nonlinear composites. Numerous studies dealt 
with nonlinear cell calculations by the Finite Element Method (FEM) (see for example
{  Adams} and 
{  Donner} \cite{ADA67}, {  Christman} {\it et al} \cite{CHR89}, 
{  Tvergaard} \cite{TVE90},  {  Michel} and {  Suquet} \cite{MIC93}).
Most of them are
limited to ``simple" microstructures, one or two inclusions embedded in a
volume of matrix.   
The need to incorporate more detailed information on the
microstructure is clearly recognized.
Recently, several studies have considered 
``complex" microstructures involving a significant number of inclusions with
irregular shape. 
{  Brockenborough}  {\it et al} \cite{BRO91}, 
{  B\"{o}hm} {\it et al} \cite{BOH93}, 
{  Nakamura} and {  Suresh} \cite{NAK93}, 
{  Dietrich} {\it et al} \cite{DIE93}, {  Becker} and {  Richmond}
\cite{BEC94} are some of the contributions to this recently developed subject.
All were based on the FEM. 
The difficulties due to meshing and to the large number of degrees of freedom 
 required by the analysis limit the complexity of the 
microstructures which can be investigated by this method. 

A typical example of a complex microstructure which is difficult to mesh and
therefore to handle by means of the FEM is shown in Figure
\ref{bornert} taken from the work of 
{  Bornert} \cite{BOR96}. 
The digital image of this Iron/Silver blend 
was obtained by Scanning Electron Microscopy (SEM). 
The initial idea of the  method proposed in \cite{MOU94} was to make direct use 
of these digital {\it images of the  real microstructure} in the numerical
simulation.
A similar
idea can be found in {  Garboczi} and {  Day} \cite{GAR95} who used a spring
network technique. 

\vskip 0.3cm
The proposed method avoids the difficulty due to meshing. It  makes use of
Fast Fourier Transforms (FFT) to solve the unit cell problem \footnote{During the revision of this
paper, the attention of the authors was called on a similar work by { 
M\"uller} \cite{MUL96} concerning phase transformation.}, even  when the
constituents have a nonlinear behavior.
FFT algorithms require data sampled in a grid of regular spacing, 
allowing the direct use of  digital images of the microstructure. 
The second difficulty (size of the problem) is partially overcome by
an iterative method not requiring the formation of a stiffness matrix. 
\vskip 0.3cm
The interest in numerical simulations of the nonlinear response of composites 
has recently been strengthened by the development of theoretical methods
which analytically predict the nonlinear overall behavior of composites 
({  Willis} \cite{WIL91}, {  Ponte Casta\~neda} \cite{PON92}, 
{  Suquet} \cite{SUQ93}). Part of the present study provides 
precise numerical results for uniaxial 
loadings which 
could serve as guidelines for theoretical predictions.
\vskip 0.3cm

The body of the method and the resulting
algorithms are presented in section 2. In section 3, the 
accuracy of the method and several numerical points are
discussed (choice of the reference
medium, spatial resolution ....). In section 4 the method is applied 
to determine the local and overall responses of composites with "random"
microstructures. 
In all the cases considered in this study the models have been limited 
to two dimensional approximations. The first reason for this approximation is
the limitation on current computational capability. The second reason is that
many microstructural observations are two dimensional.

\section{The numerical method}

\subsection{Cell problem and boundary conditions}

The overall behavior of a composite is governed by the individual
behavior of its constituents and by its microstructure. 
 Its effective response  to a prescribed path of macroscopic strains or 
stresses  may be determined numerically via the resolution of the 
so-called "local problem" on a representative
volume element (r.v.e.) $V$. In this study, the "representative" 
information on the microstructure is provided by an image (micrograph)
 of the microstructure with arbitrary complexity. The image 
contains $N$ pixels, and independent mechanical properties are 
assigned individually to each pixel. Most applications involve only a
limited number of phases, although in principle each pixel could be
considered as an individual constituent.\par
The local problem consists of equilibrium equations, constitutive equations,
and boundary and interface conditions. All different phases are assumed 
to be perfectly bonded (displacements and tractions are continuous across 
interfaces). Displacements and tractions along the boundary of the r.v.e. 
are left undetermined and the local problem is ill-posed. We choose to close 
the problem with periodic boundary conditions which can be expressed 
as follows.
The local strain field $\varepsilonbf({\bf u(x)})$ is split into its 
average $\bf E$ and a fluctuation term 
$\bf \varepsilonbf({\bf u^\ast(x)})$:
$$ \bf \varepsilonbf(u(x)) \ = \ \varepsilonbf(u^\ast(x)) + 
E\quad {\rm or\ equivalently}\quad 
\bf u(x) = u^\ast(x) + E.x. $$
By assuming periodic boundary conditions it is assumed that the 
fluctuating term $\bf u^{\ast}$ is periodic (notation: $\bf u^{\ast} \ \#$), 
and that the traction  $\sigmabf.{\bf n} $ is anti-periodic in order to meet 
the equilibrium equations on the boundary between two neighboring cells 
(notation: $\sigmabf.{\bf n} \ -\# $). This local problem could be solved by
means of the FEM ({  Suquet} \cite{SUQ87}, {  Guedes} and {  Kikuchi}
\cite{GUE90}). We propose an alternate method of resolution.

\subsection{An auxiliary problem} 

First we consider the preliminary problem of a homogeneous linear elastic body
with stiffness ${\bf c}^0$ subjected to a  polarization field  $\bf
\taubf(x)$. 
\begin{equation}
\left.
\begin{array}{rcl}
\sigmabf({\bf x})& = & {\bf c}^0: \varepsilonbf({\bf u}^\ast({\bf x})) 
\ + \ \taubf({\bf x}) 
\ \ \ \forall {\bf x} \in V 
\\  \\
{\bf div}\ \sigmabf({\bf x})& = & {\bf 0} \quad 
\forall {\bf x} \in V,
\quad {\bf u}^\ast \ \#, \ \sigmabf.{\bf n} \ -\# 
\end{array}
\right\} 
\label{eq1}
\end{equation}
The solution of (\ref{eq1}) can be expressed in real and Fourier spaces, respectively, by
means of the periodic Green operator ${\bf \Gamma}^0$ associated with ${\bf c}^0$:
\beg
\varepsilonbf({\bf u}^*({\bf x})) = - {\bf \Gamma}^0 \ast \taubf({\bf x})  \quad  
\forall {\bf x} \in V,
\label{eq2}
\en
{\rm or}
\beg
{\hat \varepsilonbf}(\xibf)  = - {\hat {\bf \Gamma}^0}(\xibf):{\hat \taubf}(\xibf) \ \
\forall \xibf \ne {\bf 0}, \ {\hat \varepsilonbf}({\bf 0}) = {\bf 0} 
\label{eq3}
\en
The operator ${\bf \Gamma}^{0}$
is explicitly known in Fourier space (see appendix \ref{appendix1}).  When the reference material is
isotropic (with Lam\'e coefficients $\lambda^0$ et $\mu^0$) it takes the form~:
\beg 
{\hat {\Gamma}}^{0}_{ijkh}(\xibf) =  { 1 \over 4 \mu^0 \vert \xibf \vert ^2
} (
\delta_{ki} \xi_{h} \xi_j +
\delta_{hi} \xi_{k} \xi_j +
\delta_{kj} \xi_{h} \xi_i +
\delta_{hj} \xi_{k} \xi_i ) - { \lambda^0 + \mu^0  \over  \mu^0 (\lambda^0 +
2\mu^0) }
{ \xi_i \xi_j \xi_k \xi_h  \over  \vert \xibf \vert ^4 }.
\label{eq4}
\en

\subsection {The periodic Lippman-Schwinger equation}

The auxiliary problem can be used to solve the problem of an
inhomogeneous elastic composite material with stiffness $\bf c(x)$ at point
$\bf x$ under prescribed strain $\bf E$~: 
\begin{equation}
\left.
\begin{array}{rcl}
\sigmabf({\bf x}) &= & {\bf c}({\bf x}): \big( \varepsilonbf({\bf u}^\ast({\bf
x})) \ + {\bf E} \big) 
 \ \ \ \forall {\bf x} \in V \\ \\
{\bf div} \sigmabf({\bf x})& = &{\bf 0} \quad 
\forall {\bf x} \in V,\quad {\bf u}^\ast \ \#, \ \sigmabf.{\bf n} \ -\# 
\end{array}
\right\} 
\label{eq5}
\end{equation}
For simplicity $\bf E$ is assumed to be 
prescribed, although other average conditions could be considered as well 
(see appendix \ref{appendix3} for prescribed stresses).
A homogeneous reference material with elastic stiffness ${\bf c}^0$ is 
introduced and a polarization tensor
$\taubf(\bfx)$, which is unknown {\it a priori}, is defined as~:
\begin{equation}
\taubf({\bf x}) =  \deltabf  {\bf c}({\bf x}) :\varepsilonbf ({\bf u} ({\bf x})),
\quad \deltabf  {\bf c}({\bf x}) \ = \ {\bf c} ({\bf x})- {\bf c}^0.
\label{eq6}
\end{equation}
Thus, the problem reduces to the {\sl periodic Lippmann-Schwinger equation}
({  Kr\"oner} \cite{KRO72}), which reads, in real space and Fourier space respectively:

\begin{equation}
\left.
\begin{array}{l}
{\bf \varepsilonbf(u(x))}= - {\bf \Gamma}^{0}({\bf x})
\ast {\taubf}({\bf x})+ {\bf E}, \\ \\
{\widehat {\bf\varepsilonbf}}(\xibf) = 
- {\widehat {\bf \Gamma}}^{0}(\xibf) : {\widehat {\bf \taubf}}(\xibf) \quad
\forall \xibf \ne {\bf 0},\quad 
{\widehat {\bf \varepsilonbf}}({\bf 0}) ={\bf E}
\end{array}
\right\}
\label{eq7}
\end{equation}
where $\taubf$ is given by (\ref{eq6}). The Lippman-Schwinger equation is an
integral equation for $\varepsilonbf({\bf u}^*)$.

\subsection{The algorithm}
\subsubsection{Continuous algorithm}
\noindent The principle of the algorithm is to use alternately   (\ref{eq6}) 
and (\ref{eq7}), 
in real space and Fourier space, respectively, 
in an iterative scheme, to solve (\ref{eq5}):

\begin{equation}
\left.
\begin{array}{rl}
{Initialization:}\ \ \ \ \ \
&{\bf \varepsilonbf}^0({\bf x}) = {\bf E},\quad \forall \ {\bf x} \in \ V,\\ 
&{\sigmabf}^{\rm 0}({\bf x}) = 
{\bf c}({\bf x}):{\varepsilonbf}^{\rm 0}({\bf x}),
\quad \forall \ {\bf x} \in \ V,  \\  \\
Iterate\ {\rm i+1}:\ \ \ \ \ \
&{\varepsilonbf}^{\rm i} \ {\rm and} \ \sigmabf^{\rm i} \ {\rm being \ known} \\
a)\ \ &{\taubf}^{\rm i}{\bf (x)} =
{\bf \sigmabf}^{\rm i}{\bf (x)}-{\bf c}^0:
{\varepsilonbf}^{\rm i}{\bf (x)},\\
b)\ \ &{\widehat {\bf \taubf}} ^{\rm i} = {\cal F}({\taubf} ^{\rm i}),\\
c) \ \ &{\rm Convergence \ test }, \\
d)\ \ &{\widehat {\varepsilonbf}} ^{\rm i+1}(\xibf) =
- {\widehat {\bf \Gamma}}^{0}(\xibf) : {\widehat {\taubf}} ^{\rm i}(\xibf)
\ \forall \xibf \ne {\bf 0}
\ {\rm and}
\ \widehat {\bf \varepsilonbf} ^{\rm i+1}({\bf 0}) = {\bf E},\\
e)\ \ &{\bf \varepsilonbf} ^{\rm i+1} = {\cal F}^{-1}(\widehat {\varepsilonbf}
^{\rm i}), \\
f)\ \ &{\bf \sigmabf}^{\rm i+1}(\bf x) = 
{\bf c}({\bf x}):\varepsilonbf^{\rm i+1}({\bf x}). 
\end{array}
\right\} 
\label{alg1}
\end{equation}
$\cal F$ and ${\cal F}^{-1}$ denote the Fourier transform and the inverse
Fourier transform. This algorithm can be further simplified by noting that 
$$
{\bf \Gamma}^0 \ \ast \ ({\bf c}^0:\varepsilonbf) = 
\varepsilonbf.$$
The modified algorithm reads~:
\beg
\left. 
\begin{array}{rl}
{Initialization:} \ \ \ \ \ \
&{\varepsilonbf}^{\rm 0}({\bf x}) = {\bf E},
\quad \forall \ {\bf x} \in \ V,\\
&{\sigmabf}^{\rm 0}({\bf x}) = 
{\bf c}({\bf x}):{\varepsilonbf}^{\rm 0}({\bf x}),
\quad \forall \ {\bf x} \in \ V,  \\  \\
Iterate\ {\rm i+1}:\ \ \ \ \ \ 
&{\varepsilonbf}^{\rm i} \ {\rm and} \ \sigmabf^{\rm i} \ {\rm being \ known} \\
a)\ \ &{\hat {\bf \sigmabf}} ^{\rm i} = {\cal F}({\sigmabf} ^{\rm i}),\\
b) \ \ &{\rm Convergence\ test}, \\ 
c)\ \ &{\hat {\varepsilonbf}} ^{\rm i+1}(\xibf) = 
{\hat \varepsilonbf}^{\rm i}(\xibf)
- {\hat {\bf \Gamma}}^{0}(\xibf) : {\hat {\sigmabf}} ^{\rm i}(\xibf)
\ \forall \xibf \ne {\bf 0} 
\ {\rm and} 
\ \hat {\bf \varepsilonbf} ^{\rm i+1}({\bf 0}) = {\bf E},\\
d)\ \ &{\bf \varepsilonbf} ^{\rm i+1} = 
{\cal F}^{-1}(\hat {\varepsilonbf} ^{\rm i+1}) \\
e)\ \ &{\sigmabf}^{\rm i+1}({\bf x}) = 
{\bf c}({\bf x}):{\varepsilonbf}^{\rm i+1}({\bf x}),
\quad \forall \ {\bf x} \in \ V, \\
\end{array}
\right\} 
\label{alg2}
\en
Convergence is reached when  
${\sigmabf}^{\rm i+1} $ is in equilibrium. The error serving to check convergence
is~:
$$
 e^{\rm i} = 
{ \left(< \vert \vert {\rm div}({\bf \sigmabf}^{\rm i}) \vert
\vert^2 > \right)^{1/2}
\over \vert \vert <\sigmabf ^{\rm i}> \vert \vert }={ \left( < \vert \vert
{\xibf}.\hat {\sigmabf}^{\rm i}({\xibf}) \vert \vert^2 > \right)^{1/2}
\over \vert \vert {\hat {\sigmabf}} ^{\rm i}({\bf 0})  \vert \vert } .
$$
The iterative 
procedure is stopped  when the error $e$  is smaller than a prescribed value 
(typically $10^{-4}$ in our calculations).

\subsubsection{Discrete algorithm}

The unit cell is discretized into a regular grid consisting of $N_1 \times N_2$ pixels (two-dimensional problem), 
or $N_1 \times N_2 \times N_3$ "voxels" 
(tri-dimensional problem). The data and the unknowns used in the numerical calculations 
are images sampled on this grid ($N_1 \times N_2$ or $N_1 \times N_2 \times N_3$ arrays). 
In two dimensions, the coordinates of the pixel  
labeled by $i_1, i_2$ are
$$ {\bf x}_d(i_1,i_2) = \left( 
(i_1-1) \cdot {T_1 \over N_1}, 
(i_2-1) \cdot {T_2 \over N_2} 
\right),\quad i_1=1,...N_1,\quad i_2=1,...N_2,
$$
where $T_j$ is the period of the unit cell in $j^{th}$ direction 
($j = 1, 2$). 
This discretization is classical in image processing. Images of 
microstructures, obtained for instance by S.E.M. (scanning electron microscopy),
 can therefore be directly used in calculations without any operation by the 
user (meshing or interpolation). This discretization is also appropriate 
for using Fast 
Fourier Transforms (FFT) packages, which contribute significantly to the 
performances of the method.
\medskip
The continuous algorithm (\ref{alg2}) has been implemented in the following  
discrete form :
\beg
\left. 
\begin{array}{rl}
{Initialization:}
&\quad {\varepsilonbf}^{\rm 0}({\bf x}_d) = {\bf E},
\quad \forall \ {\bf x}_d \in \ V, \ \\ 
 &\quad {\sigmabf}^{\rm 0}({\bf x}_d) = 
{\bf c}({\bf x}_d):{\varepsilonbf}^{\rm 0}({\bf x}_d),
\quad \forall \ {\bf x}_d \in \ V, \ \\  \\
Iterate\ {\rm i+1}:
&{\varepsilonbf}^{\rm i} \ {\rm and} \ 
{\sigmabf}^{\rm i} \ {\rm known\ at\ every }\ {\bf x}_d \\
a)\ \ &{\hat {\bf \sigmabf}} ^{\rm i} = {\cal FFT}({\sigmabf} ^{\rm i}),\\
b) \ \ &{\rm Convergence\ test},   \\ 
c)\ \ &{\hat {\varepsilonbf}} ^{\rm i+1}(\xibf_d) = 
{\hat {\varepsilonbf}} ^{\rm i}(\xibf_d)
- {\hat {\bf \Gamma}}^{0}(\xibf_d) : {\hat {\sigmabf}} ^{\rm i}(\xibf_d)
\ \forall \xibf_d \ne {\bf 0} 
\ , 
\ \hat {\bf \varepsilonbf} ^{\rm i+1}({\bf 0}) = {\bf E},\\
d)\ \ &{\bf \varepsilonbf} ^{\rm i+1} = 
{\cal FFT}^{-1}(\hat {\varepsilonbf} ^{\rm i}) \\
e)\ \ &{\sigmabf}^{\rm i+1}({\bf x}_d) = 
{\bf c}({\bf x}_d):{\varepsilonbf}^{\rm i+1}({\bf x}_d),
\quad \forall \ {\bf x}_d \in \ V \\
\end{array} 
\right\} 
\label{alg3}
\en
where ${\bf x}_d$ denote the coordinates of pixels in real space, 
and $\xibf_d$ denote the $N_1 \times N_2$ corresponding frequencies in Fourier space. 
To be more specific, the discrete frequencies are (in dimension 2) when $N_j$ is even~:
$$ 
\xi_j = 
(-{N_j \over 2}+1) \ {1 \over T_j}, 
\ (-{N_j \over 2}+2) \ {1 \over T_j}, 
\ ..., 
\ -{1 \over T_j}, \ 0, \ {1 \over T_j}, 
\ ..., 
({N_j \over 2}-1) \ {1 \over T_j}, \ {N_j \over 2} \ {1 \over T_j}, 
$$
and when $N_j$ is odd~:
$$ 
\xi_j = 
-{N_j -1 \over 2} \ {1 \over T_j}, 
\ ..., 
\ -{1 \over T_j}, \ 0, \ {1 \over T_j}, 
\ ..., 
\ {N_j -1 \over 2} \ {1 \over T_j}. 
$$

The discrete error serving to check convergence is~:
$$ e^{\rm i} = \ 
{ \left(\displaystyle  { 1 \o N } \sum_d \vert \vert 
\xibf_d \cdot {\hat \sigmabf}^{\rm i}({\xibf_d}) \vert \vert ^2 \right)^{1/2}
\over
\vert \vert\displaystyle  \hat{\sigmabf}^{\rm i}(\bf 0) \vert \vert
}
$$
(where $N=N_1 \times N_2$ is the total number of pixels).

When the spatial resolution is low and when the number $N_j$ of 
discretization point is even, a special attention must be paid to the
{ highest frequency
  \footnote{An error had crept into the expression of the highest frequency
    in the original paper published in Comput Methods Appl Mech Eng.
    The authors thank Anthony Rollett for pointing it out.
    HM 11/12/2020
    }
  $\xi_j = \pm \left({N_j \over 2} \right) {1\over T_j}$, 
  $j=1$ or $2$.
}
In most FFT packages, the Fourier expansion at these frequencies 
consists of either $\cos(\xi_j x_j)$ or $\exp(-{\rm i}\xi_j x_j)$, instead of the
correct expression consisting of the two terms 
$\exp(-{\rm i}\xi_j x_j)$ {\it and} $\exp({\rm i}\xi_j x_j)$. Therefore,
 even when the stress $\bfs$
is correctly approached by its Fourier expansion in step a) of the
algorithm (10), 
the result of step d) may not approach accurately the Fourier expansion of
the strain $\varepsilonbf$ at these particular frequencies. This is because 
$\hat{\bf {\Gamma}}^0$ is neither  even nor odd with respect to each individual
component $\xi_j$. Oscillations were observed when (4) was used with 
relatively small values of $N_j$ (lower than 128).
This problem was fixed by using a different expression of
$\hat{\bf {\Gamma}}^0$ in algorithm (10) at these frequencies
$$\hat{\bf {\Gamma}}^0 = \left({\bf c}^0 \right)^{-1}.$$
In other terms,  the stress $\bfs$ is forced to $\bf 0$  by the algorithm 
at these frequencies when convergence is reached.

\subsection {Nonlinear Behavior.}

\begin{figure} 
\begin{center} {
\leavevmode
\includegraphics[width=10cm]{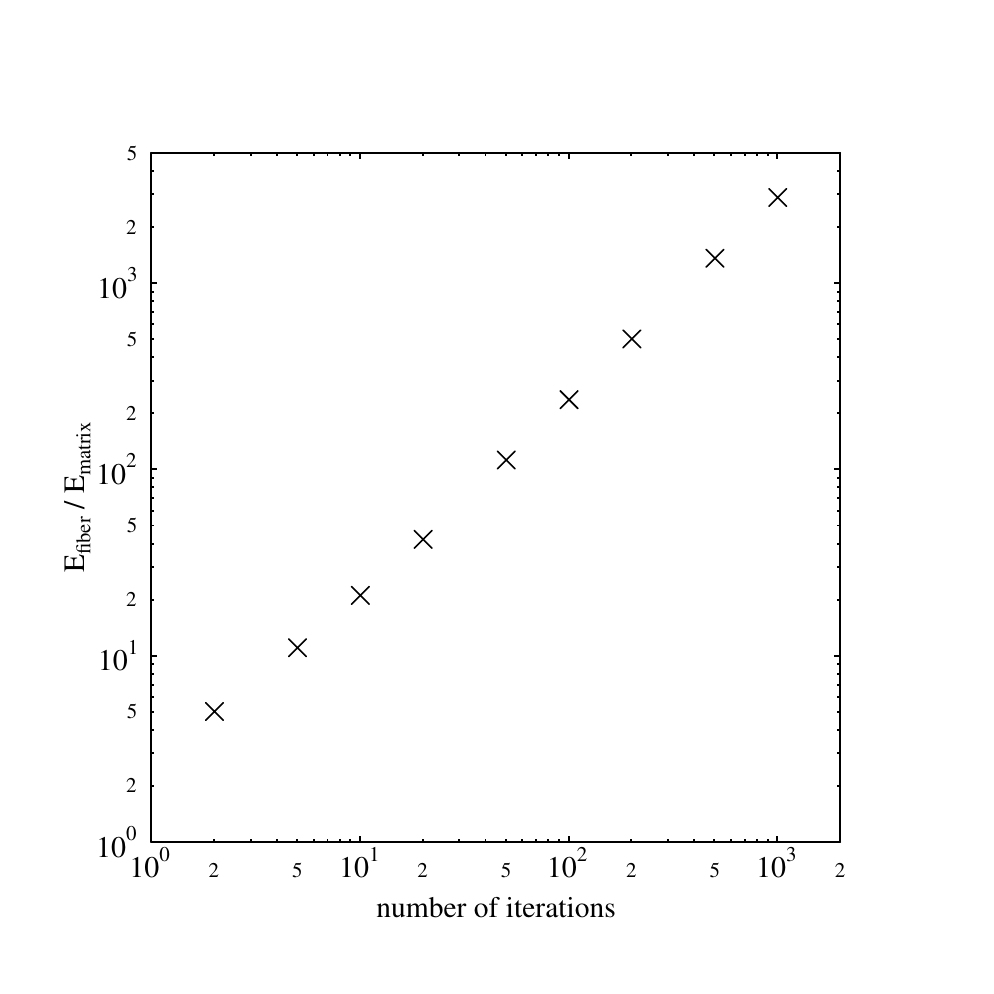}}
\end{center}
\caption{
{Transverse Young's modulus of the composite. Dependence of the number of iterations at 
convergence on the contrast of the elastic moduli of phases  ($e \leq
10^{-4}$). 
Square array as shown in Figure \ref{std}. Spatial
resolution $128 \times 128 $ pixels. Fiber volume fraction 47.5 \%. Poisson
coefficients $\nu^f=\nu^m=0.35$. The stiff phase is the fiber. 
}
} 
\label{fig1}
\end{figure}

\begin{figure} 
\begin{center} {
\leavevmode
\includegraphics[width=10cm]{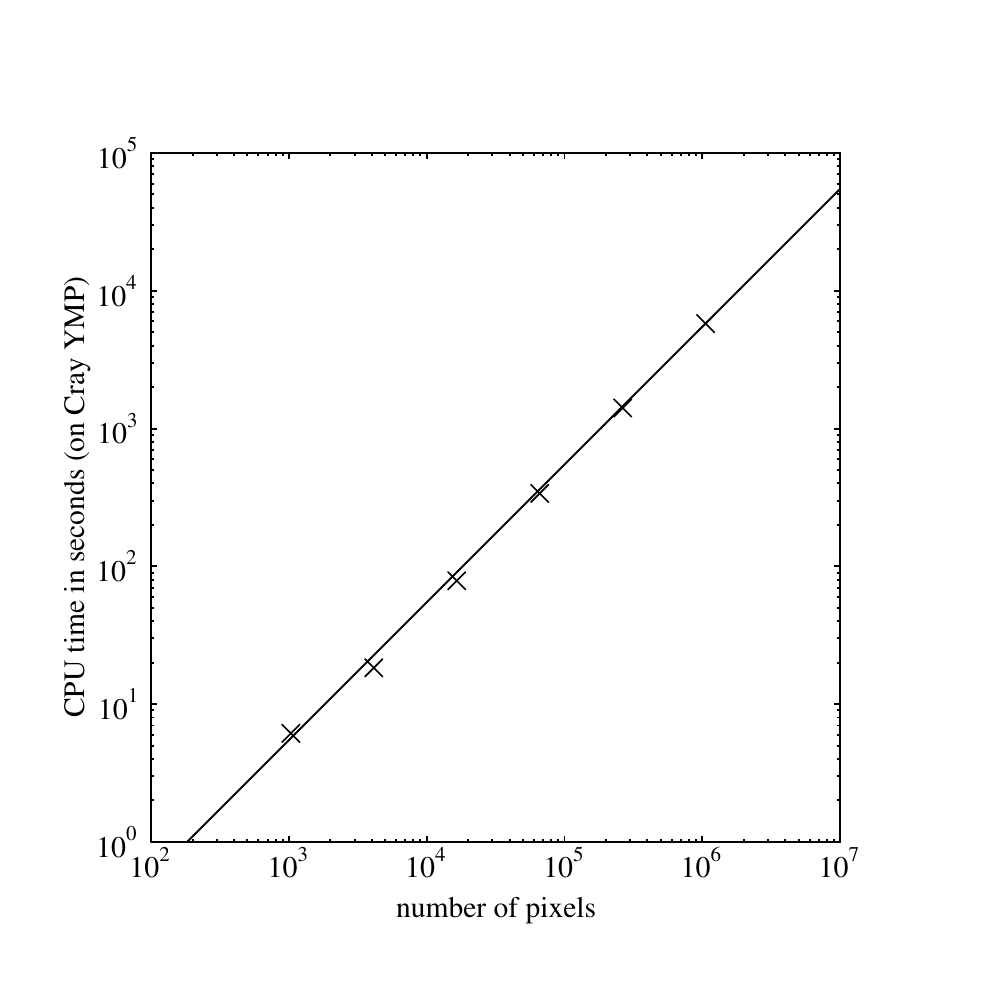}
}
\end{center}
\caption{{CPU time on one processor of a CRAY YMP as a function of the size $N$ of the
problem. The solid line is obtained by linear regression on all points and
passes through the origin. 
}}
\label{fig2}
\end{figure}

\noindent The algorithm can be extended to the case in which the
 individual constituents obey a nonlinear law, written either in terms of
stresses and strains (nonlinear elasticity at infinitesimal strain) or in
incremental form relating strain-rates and stress-rates (flow theory).
The nonlinearity requires an appropriate modification of 
step {\it e}) in algorithm (\ref{alg3}).
In the present study, special attention will be paid to phases exhibiting an
incremental 
elastic-plastic behavior at small strains governed by a $J_2$-flow theory 
with isotropic hardening (although more general 
constitutive laws can be considered)~:
\beg
{\dot \sigmabf} = {\bf c}:({\dot \varepsilonbf}-{\dot 
\varepsilonbf}^{p}),\quad \dot {\varepsilonbf}^{p} = {\dot p}\ {3 \over 2} {{\bf 
s} \over  {\sigma_{eq}}},\quad {\sigma_{eq}}-{\sigma_{0}}(p) \leq 0,\quad  {\dot 
p} \geq 0.
\label{comp}
\en
$\varepsilonbf^{p}$ denotes the plastic strain, $\bf s$ denotes the stress
deviator and $p$ denotes the hardening 
parameter, which coincides with the cumulated plastic strain
$$
\dot{p}(t) =  \left( {2 \over 3 } \dot{\varepsilon}^{p}_{ij}(t)
\dot{\varepsilon}^{p}_{ij}(t) \right),
\quad p(t) = \int_0^t \dot{p}(s) \ ds
\quad \sigma_{eq} = \bigg( \frac{3}{2} s_{ij} s_{ij} \bigg)^\frac{1}{2}
.
$$ 
\vskip 0.3cm
The integration in time of the constitutive law (\ref{comp}) is achieved by 
means of an implicit scheme which is classical in the analysis of 
elastic-plastic structures by the FEM method. 
The time interval (or, alternatively, the loading path) is 
discretized into subintervals $[t_n,t_{n+1}]$. The field equations are 
solved for $\left(\varepsilonbf_{n},
\sigmabf_{n},{p}_n \right)$, which denote  strain, stress and 
hardening parameter at time $t_n$. 
Assuming that these fields are known at step $n$ (time  $t_n$), the principal
unknown at step $n+1$ is $\varepsilonbf_{n+1}$. The incremental equations 
(\ref{comp}) are discretized by an implicit scheme. The unknown 
$\varepsilonbf_{n+1}$ is 
a compatible strain field such that the associated stress field (by the 
constitutive law) is in equilibrium. The  resulting system of equations 
to be solved for $\varepsilonbf_{n+1}$ is nonlinear. The  
algorithm for the determination of $\varepsilonbf_{n+1}$ reads 
(for simplicity the lowerscript $(n+1)$ is omitted below; 
superscripts i and i+1 refer to the iterative loop within the
step)~:
\begin{equation}
\left. 
\begin{array}{rl}
{Initialization:}  
&\varepsilonbf^{\rm 0}({\bf x}_d) \ {\rm given\ by}\ (\ref{ini}),\\
& {\rm Compute} \ \sigmabf^{\rm 0} \ {\rm and} \ {p}^{\rm 0}\ 
{\rm from} \  (\varepsilonbf^{\rm 0},
\sigmabf_{n},\varepsilonbf_{n},{p}_{n}), \\ \\
{Iterate}\  {\rm i+1}:
&{\varepsilonbf}^{\rm i} \  {\rm and} \ {\sigmabf}^{\rm i} \
{\rm are \ known} \\
a)\ \ &{\hat {\bf \sigmabf}}^{\rm i} = 
{\cal FFT} ({\sigmabf}^{\rm i}),\\ 
b) \ \ &{\rm Convergence \ test}, \\
c)\ \ &{\hat {\varepsilonbf}}^{\rm i+1}(\xibf_d) = 
{\hat \varepsilonbf}^{\rm i}(\xibf_d)
- {\hat {\bf \Gamma}}^{0}(\xibf_d) : {\hat {\sigmabf}} ^{\rm i}(\xibf_d)
\ \forall \xibf_d \ne {\bf 0}, 
\ \hat {\bf \varepsilonbf}^{\rm i+1}({\bf 0}) = {\bf E}_{n+1}, \\
d)\ \ &{\bf \varepsilonbf}^{\rm i+1} = 
{\cal FFT}^{-1}(\hat {\varepsilonbf}^{\rm i+1}) \\
e)\ \ & {\rm Compute}\ {\sigmabf}^{\rm i+1} \ 
{\rm and}\ {p}^{\rm i+1}\ 
{\rm from}\  (\varepsilonbf^{\rm i+1}, 
\sigmabf_{n},\varepsilonbf_{n},{p}_{n})\\ 
\end{array}
\right\} 
\label{alg4}
\en

More specifically
\begin{itemize}
\item[a)]
The initial strain  
$\varepsilonbf^{\rm 0}$ at time $t_{n+1}$ is extrapolated (linearly) from
$\varepsilonbf_{n}$ and 
$\varepsilonbf_{n-1}$ at the two previous time steps $t_n$ and $t_{n-1}$~:
\beg
\varepsilonbf^{\rm 0}({\bf x}_d) = 
{\varepsilonbf}_{n}({\bf x}_d) +
{t_{n+1}-t_n \over t_n-t_{n-1}}
(\varepsilonbf_{n}({\bf x}_d) - \varepsilonbf_{n-1}({\bf x}_d)),
\quad \forall \ {\bf x}_d \in \ V.
\label{ini}
\en
This choice significantly improves the convergence of the
iterative process within the time step.
\item[b)]
${\sigmabf}^{\rm i}$ and ${p}^{\rm i}$  are computed from 
$(\varepsilonbf^{\rm i},\sigmabf_{n},\varepsilonbf_{n},
{p}_{n})$ (step {\it e}) in algorithm (\ref{alg4})) by a radial 
return method  (see appendix \ref{appendix2}). 
\end{itemize}

\section{Convergence and accuracy of the method}

\subsection{Reference medium}

The rate of convergence of the algorithm depends drastically on 
the Lam\'e coefficients $\lambda^0$ and $\mu^0$ of the reference material. 
After several tests, the best rate of convergence was observed with
\beg
\left.
\begin{array}{rl}
\lambda^0 &= {1 \over 2} \left( 
\displaystyle \doublelow{{\rm inf}\cr {\bf x} \in V \cr} \lambda({\bf x}) + 
\doublelow{{\rm sup}\cr {\bf x} \in V \cr}
 \lambda({\bf x}) \right) \\ \\
\mu^0 &= {1 \over 2} \left( \displaystyle
\doublelow{{\rm inf}\cr {\bf x} \in V \cr} \mu({\bf x}) + 
\doublelow{{\rm sup}\cr {\bf x} \in V \cr} \mu({\bf x}) \right)
\end{array}
\right\}
\label{conv}
\en
The number of iterations at convergence is significantly influenced by 
several other parameters. First, 
as shown in Figure \ref{fig1}, it increases with the contrast between the phases 
(typically the ratio between the elastic moduli of the phases). When the 
contrast is infinite (rigid inclusions or voids in an elastic matrix), 
the algorithm  no longer converges. 
Second, the number of iterations at convergence also depends on the complexity of 
the solution itself. In the example of an elastic ideally plastic matrix 
reinforced by stiff inclusions, the computing time increases with the 
tortuosity of the bands where the strain tends to localize (see below).

\subsection{Implementation of the method on a vector or a parallel
computer}

The constitutive law acts locally in real space ({\it i.e.} applies 
separately to each individual  point $\bfx$). Similarly, 
Green's function ${\bf \Gamma}^0$ acts locally   in Fourier space,
({\it i.e.} applies separately to each individual frequency $\xibf$). From a
computational standpoint, the corresponding steps (c and e in the
algorithms (\ref{alg3}) or (\ref{alg4}) ) are performed by
independent loops on
each individual pixel in real or Fourier space. These steps can consequently be vectorized or 
parallelized. In addition, optimized FFT packages are available on
most vector or
parallel computers. The whole algorithm can therefore be efficiently 
implemented on
these machines. 

It follows from the  same argument that the time spent in the steps
corresponding to the constitutive law and to the Lippman Schwinger equation
varies linearly with the number $N$ of pixels. The CPU time for a  
FFT varies as
$N \cdot \log _2N$. The time required by the other steps of the algorithm
are comparable to the time required by the FFTs. The CPU time $t$ for one
iteration can be estimated by
$$ k_1 \times N \ \ \le \ \ t \ \ \le \ \ k_2 \times \ N \log _2N, $$
where $k_1$ and $k_2$ are expected to be independent of the size $N$ of the
problem.
The dependence of the CPU time on the size of the problem is shown in
Figure \ref{fig2}. The square unit cell shown in Figure \ref{std} is subjected to uniaxial 
transverse tension at $0^0$. The 
volume fraction of fibers is 47.5\%. Both the fibers and the matrix are
assumed to be elastic with elastic constants given by (\ref{fibr}) and
(\ref{matr0}). The dependence of the CPU time on the size of 
the problem is approximately linear.
\vskip 0.3cm
\noindent
{\bf Optimizing the memory occupancy}.
The Fourier transform of a real valued function has the symmetry property
$$\hat{f}(-\xibf) = \overline{\hat{f}}(\xibf).$$


Since all quantities under consideration in our computation are real, this
symmetry property allows us to restrict our attention to positive frequencies (the
values of the fields for negative frequencies being immediately deduced).
The size of the arrays can therefore be divided by 2, provided  the FFT
package allows for the storage of real numbers as complex numbers with the
same memory occupancy.

\vskip 0.3cm
\noindent {\bf Performances}.
Most computations were run on a Cray YMP with  peak performance
of $333 \ MFlops$. The performance observed with our algorithm was 
$\simeq 210 \ MFlops$ on the elastic-plastic problem described in section 4 with
unit cells discretized into 
$1024 \times 1024 \ pixels$.
The typical CPU time on one processor of this computer is less than 30
seconds for an elastic problem (with a spatial resolution of 
$1024\times 1024 \ pixels$, the ratio between the Young moduli being
approximately  6). When the matrix is elastic plastic, the typical CPU time
for a run as described in section 4 is 4000 seconds.

\subsection{Comparison with analytical solutions}

To assess the accuracy and the stability of the method we examined two cases 
for which analytical solutions are available. 
\vskip 0.3cm
\noindent
{\bf Laminates}. 
The first example concerns 
layered materials. As is well-known, the strain field is then uniform within 
each individual layer and takes different values from one layer to another. 
The example shown in Figure \ref{layer} corresponds to a two-phase 
material, both phases having equal volume fraction. The layers are parallel 
to the  plane $(x_2,x_3)$. 
The constitutive materials of the layers were linear elastic with elastic 
characteristic given by (\ref{fibr}) and (\ref{matr0}). 
The applied loading was pure shear parallel to the layers 
$$
\S_{12} \ {\rm arbitrary},\quad \S_{11}=\S_{22}= \S_{33} = 
\S_{13}= \S_{23}= 0.
$$
The image was discretized into $32 \times 32$ pixels (good results were
obtained with an even cruder resolution). The computed local strain field 
$\varepsilon_{12}$ is plotted in Figure \ref{layer} and shows no oscillation.
In addition the numerical solution coincides with the
exact solution. 
\vskip 0.3cm
\noindent
{\bf Circular fiber at dilute concentration}. 
The second example concerns the elastic strain field 
generated by stiff circular fibers placed at the nodes of a square 
lattice in a more compliant matrix. 
The exact solution to this problem (with periodic boundary conditions) 
is not known in closed form (to the authors' knowledge). However when the 
volume fraction of fibers is small 
this solution can be accurately approximated by the solution of a simpler problem,
where a circular fiber (with radius $a$) is surrounded by a circular shell of
matrix (with radius $b$) and subject to the boundary condition
$$ {\bf u}(\bfx) = \bfE.\bfx \quad {\rm when}\quad r=b,$$
where the overall strain $\bfE$ is the same as in the original periodic 
problem. When the imposed loading is an in-plane shear
$E_{12} \neq 0$, other $E_{ij}=0$, the displacement field has the form
$$
\left.
\begin{array}{rl}
\displaystyle u_r(r,\theta) & = \left( A r^3 + B r + {\displaystyle C \o \displaystyle r} + 
{\displaystyle D \o \displaystyle r^3} \right) {\rm
sin}(2 \theta),
\\ \\
\displaystyle u_{\theta}(r,\theta)& = \left( {\displaystyle 2 \l + 3 \m \o \displaystyle \l} 
A r^3 + B r + {\displaystyle \m \o \displaystyle \l + 2 \m}{\displaystyle C \o \displaystyle r} 
- {\displaystyle D \o \displaystyle  r^3} \right) {\rm cos}(2 \theta),
\end{array}
\right\}
$$
where $r$ and $\theta$ are the polar coordinates in the plane. 
$A,\ B,\ C,\ D,$ take different values in the matrix and in the fiber. They
solve a system of linear equations expressing the boundary condition at
$r=b$, the absence of singularity at $r=0$, the continuity of tractions and
displacements at $r=a$. 
\vskip 0.3cm
According to Saint Venant's principle, the 
local strain fields in the two problems coincide far from the
boundary of the cell.
Therefore at low volume fraction of fibers ($a^2 / b^2 \ll 1$), 
the solutions of the two problems are expected to coincide except in the 
vicinity of the boundary of the cell. The example presented in Figure
\ref{infini} corresponds to $a/b = 1/16$. The spatial discretization used in the
numerical calculation was $1024 \times 1024$.   
The component $\varepsilon_{12}$ of the 
strain field in a square window of width $c= 4 a$ is shown in Figure
\ref{infini} (note that the unit cell itself with width $2b$ is much larger 
than the window shown). 
There is almost no difference between the  analytical and the numerical 
solutions shown in (a) and (b) respectively. A more explicit comparison 
is made in Figure \ref{infini} (c) which shows an horizontal cut through 
the  field $\varepsilon_{12}$ at $x_2 = 0$. Except from little
undulations inside the inclusion, there is no significant oscillations at  
the fiber boundary where the field  $\varepsilon_{12}$ is discontinuous. 
In addition the accuracy of the numerical solution is observed to 
increase with the spatial resolution.
The discrepancy between the numerical and the analytical solutions 
depends on the spatial resolution and should not be attributed to
a Gibbs phenomenon, {\it i.e.} to an oscillation of the Fourier series of a
function in the vicinity of a discontinuity point.  This oscillation
is attached to the summation of the Fourier series which is {\it not} what the 
discrete inverse Fourier transform performs.   
\vskip 0.3cm
\noindent{\bf Discrete Fourier transform}. 
The discrete Fourier transform, when applied to an image discretized into 
$N_1\times N_2$ pixels, is the exact Fourier transform of the image when  
two  requirements are met~: 
({  Brault} and {  White} \cite{BRA71})
\begin{itemize}
\item[\bf C1] the image is periodic with the same period $(T_1,T_2)$ as 
the unit cell,
\item[\bf C2] the image cut-off frequency 
$f^c$ ( {\it i.e.} the frequency above which the Fourier transform of the image vanishes identically) is less than half of the sampling frequency (Shannon's theorem): 
$$ f^c_j < {1 \over 2} \ {N_j \over T_j} \quad j = 1,2$$
\end{itemize}
The periodic boundary conditions which have been assumed from the true 
beginning of this study ensure that condition (C1) is met. 
However, condition (C2) is not 
met in general. In particular a discontinuous field has no cut-off frequency 
and there is no discretization able to capture this discontinuity. 
It is however expected that the solution of the discrete problem approaches
 the solution of the continuous problem when the image sampling 
(number of pixels) increases. A  high resolution will therefore be required 
for problems in which high strain or stress gradients are likely to occur. 

\subsection{Influence of spatial resolution}
As already stated the influence of the spatial resolution depends on the
stress and strain gradients within the phases and therefore on the 
strength of the phases nonlinearities. 
The following examples illustrate these general considerations.
The method has been applied to simulate the local and overall response 
of composites reinforced by unidirectional 
long fibers aligned along the $e_3$ direction. 
The geometry of these composites is described by a two-dimensional image of
their cross section. Generalized plane strains were assumed~:
\beg
{u}_1 (\bfx)={u}_1 (x_1,x_2),\quad 
{u}_2 (\bfx)={u}_2 (x_1,x_2),\quad 
{u}_3(\bfx)= E_{33} x_3.
\label{gps}
\en 
The overall strain ${\bf E}$ has four independent components $E_{11}$, $E_{22}$,
$E_{12}$, $E_{33}$ (the other two are equal to 0). The overall
stress $\bfS$ also has four independent components. 
It is possible to prescribe either a path in the space of strains, 
or a path in the space of
stresses, or alternatively some components of the strain and the other
components of the stress. Classical plane strains are a particular case of
the more general setting considered in (\ref{gps}). It corresponds to
a path in the space of strains along which $E_{33}$ is identically $0$. The
need to introduce generalized plane strain is illustrated by    
uniaxial tension in the $0^0$ direction, which corresponds to a path in the 
space of stresses along which 
\beg
 \S_{11} \ {\rm arbitrary},\quad \S_{22}=\S_{12}= \S_{33}= 0.
\label{ut}
\en
The axial component $E_{33}$ of the strain  is
unknown and determined {\it a posteriori} by the condition $\S_{33}=0$. 
The assumption of generalized plane strains reduces (\ref{eq5}) to a 
two-dimensional problem for the two unknowns $(u^*_1,u^*_2)$.
\vskip 0.3cm
Two classical configurations were investigated in which the fibers 
were placed at the nodes of a square or hexagonal lattice. 
The fibers were assumed to be elastic, isotropic, and
characterized by a Young modulus and a Poisson ratio~ :
\beg
E^f \ = \ 400 \ {\rm GPa},\quad \nu^f \ = \ 0.23.
\label{fibr}
\en
The fiber volume fraction was 47.5 \% (for comparison, we
chose the same volume fraction as in \cite{BOH93}). 
The behavior of the matrix was varied from linear elasticity to
elasto-plasticity with hardening so as  to study the effect of the
nonlinearity on the accuracy of the method. All the constitutive laws of the
matrix which were 
considered can be put in the incremental form (\ref{comp}). Its isotropic
elastic properties were characterized by a Young's modulus and Poisson coefficient 
\beg
E^m \ = \ 68.9 \ {\rm GPa},\quad \nu^m=0.35.
\label{matr0}
\en
The plastic properties of the matrix were governed by the Von Mises criterion
\beg
\s_{eq} \leq \s_0 + H p.
\label{matr}
\en
The  initial yield stress $\s_0$ was either infinite (pure linear elasticity) 
or given by $\sigma_0 \ = \ 68.9$ MPa. The
hardening modulus $H$ was either $0$ (perfectly plastic behavior) or 
$H= 1 \ 171$ MPa (isotropic linear hardening).
\vskip 0.3cm

The influence  of spatial resolution on the accuracy of the results was
studied.
The spatial resolution of the image is determined here through the square root 
of the total number of pixels contained in the image divided by the number 
of fibers in the image. For the square array, with $N_1 \times N_1$ pixels and a single fiber in
the unit cell, the spatial resolution is exactly $N_1$.
The hexagonal array can be viewed as a rectangular array, thus allowing 
the use of the Fourier technique
in orthogonal coordinates, instead of the natural nonorthogonal coordinates
defined by the two unit vectors of the hexagonal lattice (see Figure \ref{std}). 
The rectangular unit cell contains 
$1+4\times{1 \over 4} \ = \ 2$ fibers.
The number of
pixels along the first direction $x_1$ is 2 times larger than
the number of pixels in the second direction $x_2$. The spatial step 
in $x_2$ is $2 \sqrt{3} / 3$ times larger than the step in $x_1$.
Therefore in the hexagonal array, the spatial definition as defined above 
is again $N_1$ for an image containing $2N_1 \times N_1$ pixels.
\vskip 0.3cm
Both unit cells were submitted to uniaxial tension at $0^0$ and $45^0$ in
the sense of (\ref{ut}).
The results of the overall response of the composite are shown  
in Tables 1 to 6. The initial response of the composite is linear
and its slope  defines the overall Young's modulus of the composite.
When the matrix is elastic ideally plastic the overall stress  applied to the
composite in the direction of tension reaches (asymptotically)  a limit
which defines  the overall {\it flow stress} of the composite. 
When the matrix is governed by a linear
hardening, the stress-strain curve of the composite exhibits a nonlinear
transition to an asymptotically linear (affine) response. The slope of this 
limit response is the overall hardening modulus of the composite.
\vskip 0.3cm
Each table gives an overall material constant as a function
of the spatial resolution of the image. The "error" was estimated as the
relative difference between the result at a given resolution 
and the result at the finest resolution. 
\vskip 0.3cm
These results suggest  the following remarks.
\begin{itemize}
\item[1.] When both constituents are linearly elastic, the overall stiffness
is  not very 
sensitive to spatial resolution. Even at the lowest resolution
($32 \times 32$ pixels/fiber), the estimated error was under 1\% 
in all cases.
%
\item[2.] When the matrix is elastic plastic, the local and overall
responses are sensitive to spatial resolution. 
The strain fields exhibit a strong tendency to concentrate in thin bands. The
higher the nonlinearity, the thinner the bands. These stiff gradients in strain 
require high spatial resolution to be
correctly captured. 
\item[3.]
The solutions may even be discontinuous when the matrix is elastic-perfectly 
plastic. This explains the relatively high errors at low resolution: about 15\% for 
the square array of fibers in an elastic-perfectly plastic matrix under
tension at $0^\circ$, with a resolution of  $32 \times 32$ \ pixels/fiber.
Shear bands can form in the matrix under tension at $45^\circ$.
These shear bands correspond to a mode of deformation of the
r.v.e. in plane strains. Therefore, for this particular loading,  the effective behavior of 
the composite depends only on the behavior of the matrix.  The overall
flow stress of the composite coincides with the flow stress of the matrix under
plane strain conditions, {\it i.e.} $2 \s_0 \over \sqrt{3}$.
The formation of a slip plane through the matrix is well captured by the
numerical method and explains the precision of the numerical result for this
particular loading. 
\item[4.]
When the matrix has linear hardening, the strain fields are more regular
than in the perfectly plastic case.
The local and overall
responses of the composite are less sensitive to spatial resolution. The error on the hardening
 modulus is about 7.5\% with a resolution of  $32 \times 32$ \ 
pixels/fiber.
\end{itemize}

This study of the influence of spatial resolution led us to use a 
resolution of $128 \times 128$  pixels/fiber in most of the examples 
presented in the next section.

\section{Fiber arrangement}

In this section we investigate the influence of the geometrical arrangement of 
the fibers
on the local and overall responses of nonlinear composites. Attention is
again
restricted to two-dimensional problems, {\it i.e.} to composites reinforced by
aligned fibers. The fiber arrangement is determined by a two-dimensional
image of the composite cross section.
 
\subsection{Configurations}

Two classes of fiber arrangement, regular and random, were considered. 
The fibers were identical circular disks and they were 
not allowed to overlap (impenetrability condition) except in section 4.3. In most simulations the 
fiber volume fraction was prescribed to $47.5\%$, except in section 4.3. 
\vskip 0.3cm
\noindent
{\bf Standard fiber distribution}.
The ``standard" configurations consist of a single fiber placed at
the nodes of a square or an hexagonal lattice (see preceeding
section). 
Most F.E.M. cell calculations reported in the literature are based on these 
standard configurations with the exceptions of {  Brockenborough} {\it et al}
(1991) and {  B\"{o}hm} {\it et al} (1993)  who investigated the effect of
disorder in the fiber arrangement on the overall transverse properties of
composites.

\paragraph{Random fiber distribution}

In the "random" configurations, the centers of the fibers were placed at
random in the unit cell, subject only to the constraints of impenetrability
and periodicity. The latter constraint implies that, when a fiber overlaps 
the boundary of the unit cell, it is split into two 
parts I and II (see Figure \ref{fcar} ) to fit in the unit cell. 
The size of the images was the largest one allowed by the
memory  on our computer and compatible with a resolution of
$128 \times 128$ pixels per fiber. These two constraints led to unit cells 
discretized into $1024 \times 1024$ pixels and containing up to 64 fibers.

\subsection{Impenetrable fibers}

Twenty three different configurations of 64 impenetrable fibers
 were generated randomly in the unit cell.
The fibers were assumed to be elastic with material properties
given by (\ref{fibr}).
The matrix was an elastic plastic material governed by a $J_2$ flow theory 
(\ref{comp})
with material properties given by (\ref{matr0}) (\ref{matr}).
The local and global responses of each configuration to a transverse uniaxial 
tension in the $0^0$ direction (according
to (\ref{ut})) were computed with the above described method.
The square array and hexagonal array were also subjected
to transverse tension in the $0^0$ and $45^0$ directions. 

\subsubsection{Local and overall responses}
The stress-strain  curves predicted by the simulation are shown in Figure
\ref{courbes}. 
The solid line corresponds to the mean response (average of the
stress-strain curves over the 23 configurations).

These results call for the following comments~:\begin{itemize}
\item[1.]
The fibers were stiff and perfectly bonded to the matrix. Therefore, although
the strain $E_{33}$ in the axial direction was not imposed {\it a priori}
($\S_{33}$ was prescribed to $0$), it was relatively small along the whole 
loading path. The strain state was consequently close to the plane strain
state,
explaining the strain concentrations  observed in the 
perfectly plastic matrices. 
As is well known, plane strain is more favorable to these strain concentrations
than is pure uniaxial tension.  
 
\item[2.]
The square lattice has a marked transverse anisotropy 
which is strengthened by the nonlinear behavior, which gives raise to different 
responses when the direction of tension makes an angle of $0^0$ or $45^0$ 
with one of the axes of the square lattice. The low value of the flow stress in the 
diagonal direction ($45^0$)  is due to  a shear plane passing 
through the matrix. Indeed, when a 
plane of shear can be passed through the weakest phase of a composite, the 
shear strength of the composite is exactly the strength of the weakest phase
({  Drucker} (1959)). In 
tension (under plane strains) in a direction inclined at $45^0$ on this plane, the 
transverse flow stress of the composite is $2\s_0^m / \sqrt{3}$. 
This is the flow stress observed in Figure \ref{courbes} and Table 4 
($2\s_0^m / \sqrt{3}\simeq 79.56$ MPa). In conclusion, 
except at low volume fractions, the square array should not be used 
to investigate 
the transverse properties of transversely isotropic nonlinear composites.

\item[3.] 
The hexagonal lattice approaches transverse isotropy.  When the matrix is 
a hardening material, the predictions obtained with the hexagonal lattice 
underestimate the stiffness of the composite, or at least are located below 
the average of the predictions for the random configurations in the range of 
overall deformations considered. Another computation, not reported here, 
was performed up to 30\% of transverse strain, with no modification in
the conclusions. A similar observation was made by {  Brockenborough} 
{\it et al} (1991) for another system. When the matrix is ideally plastic, 
the low value of the flow stress in the diagonal direction ($45^0$)  is again 
due to a shear plane passing through the matrix. In conclusion, the 
hexagonal lattice should be used with care to predict the transverse 
properties of nonlinear composite systems, even for hardening matrices. 

\item[4.]
The deviation from the average 
of the transverse Young's moduli 
computed on the different configurations is small. By contrast, the 
deviations in the other properties (flow stress, hardening modulus) are 
higher and may be attributed to the combined effects of 
nonlinearity and incompressibility.

\item[5.]
The local plastic strains showed significant 
differences between the ideally plastic case and the hardening case. 
For the former, the strain concentrates in thin bands 
in the matrix. In most configurations, only a small percentage of the 
matrix contributes to the plastic dissipation. The overall flow stress 
of the composite is observed to be directly related to the "tortuosity" 
of these bands. 
Two different configurations with the
corresponding zones of strain concentration
 are shown in Figure \ref{pstrain}. In the first configuration
slip bands inclined at approximately $45^0$ on the direction of traction can
be passed  through the matrix, resulting in a low flow stress.
Conversely, the fiber arrangement in the second direction inhibits
long-range slip bands and causes these bands to deviate or the plastic
deformation to spread into wider zones. The plastic dissipation and the flow
stress are higher
in the second configuration than in the first one.
Adding more fibers in the undeformed zones
would not change the plastic dissipation, or in other terms, would not
affect the flow stress of the composite.
These results lead us to think that, when the matrix is perfectly plastic,
the geometrical parameter which governs (at first order) the flow stress of
the composite is not the volume fraction of the fibers but, instead, 
the length of the shortest path passing through the matrix at an angle 
of approximately $45^0$ in tension, or $0^0$ in shear.
 
\item[6.]
When the matrix is a hardening material, the plastic strain spreads all
over the matrix (see Figure \ref{pstrain}). The whole 
matrix contributes (although non homogeneously) to the plastic dissipation and, 
consequently, to the overall strengthening of the composite. In this case,
the volume fraction of the fibers seems to be the relevant geometrical
information (at least to first order) to predict the overall hardening of
the composite.
\item[7.]
In spite of the differences in the maps of plastic strains in the ideally
plastic material and in the hardening matrix, the "stiffest" (respectively 
the "weakest")  configurations 
in the ideally plastic case remain the stiffest (respectively the weakest) 
configurations in the hardening case.  
\end{itemize}

\subsubsection{Model size}

The present section deals with the "representativity" of a unit
cell in two aspects. First, does the unit cell contain
enough heterogeneities so that the computed effective properties no longer
depend on the cell size? Second, how much do different unit cells randomly
generated with the same volume fraction and number of heterogeneties differ
from each other?  

Several series of microstructures containing 4, 9, 16, 36, 64 or 256 
impenetrable fibers randomly placed in the unit cell were generated. 
The volume fraction of fibers was identical in all simulations ($47.5\%$)
and the spatial resolution was also fixed ($128 \times 128$ pixels/fiber). 
The total
number of pixels in each image was therefore the number of fibers multiplied
by $\times 128 \times 128$. The fibers
and the matrix
were respectively assumed to be elastic and elastic-perfectly plastic with
materials properties given by (\ref{fibr}) and (\ref{matr0}) (\ref{matr}).
The loading was uniaxial transverse tension at $0^0$ (see (\ref{ut})).
Statistical data on the computed Young's moduli as a function of the
number of fibers in the unit cell are reported in Table 7.
The mean Young's modulus and its standard deviation are defined as 
$$  {\bar E} \ \ = \ \ {1 \over N_s} \ \sum_{i=1,N_s} E_i,\quad 
\sigma(E) \ \ = \ \ \sqrt{{1 \over N_s-1} \ \sum_{i=1,N_s} (E_i-{\bar E})^2}$$
where $E_i$ is the Young's modulus of the ${\sl i}^{\rm th}$ microstructure 
and $N_s$ is the number of different microstructures.
The error on the mean is classically estimated by the ratio
$$ {\sigma(E) \over \bar{E} \sqrt{N_s}}$$
Similar data on the overall flow stress of the composite are given in 
Table 8. 
The number of fibers in the unit cell does not
significantly influence the mean overall properties, provided  a lower number 
of fibers is compensated by a higher number of configurations. The mean Young's
modulus and the mean flow stress of configurations with four fibers differ 
from those  of configurations with 256 fibers by  
$0.56 \%$ and $0.74\%$ respectively. These differences are comparable to the
error on the mean itself ($0.13\%$ and $0.23\%$ for the Young's modulus and
the flow stress for configurations with  256 fibers).
This is an illustration of the ergodic property~: spatial averaging on one 
large sample is equivalent to ensemble averaging on many small samples.  
A related observation is that the standard deviations of the overall 
properties decrease as the number of fibers increases. 

\subsubsection{Spacing between fibers.}

In the above analyses, the fibers were placed randomly 
in the unit
cell with impenetrability as the only restriction. 
The effects of imposing a minimal space between
fibers are of interest for at least two reasons. First, when the minimal
spacing between the centers of the fibers increases, the ordering of the microstructure
increases. As a limit case, when this minimal spacing 
reaches $\sqrt{{2 \over \sqrt{3}} . {S \over {N}} }$ ($S$ is 
the surface of the unit cell, N is the number of fibers), the microstructure is completely determined
and coincides with the centered hexagonal arrangement. 
Second, numerical difficulties could be expected  when two neighboring 
fibers are nearly touching. Indeed, when the spatial resolution
is not fine enough, the method cannot capture the high strain gradients in
the necks between the two fibers. 
\vskip 0.3cm
Ten configurations with 64 fibers were generated, and a minimal space 
of 4 pixels between two neighboring fibers was imposed.
This distance seemed sufficient to correctly describe strain concentration.
The results of this study suggest the following comments~:

\begin{itemize}
\item[1.]
When the matrix is elastic ideally plastic, the mean
overall flow stress is $\Sigma_0 = 86.9 $ MPa (with an estimation
error of $0.44$  MPa). This value is  $2.0\%$ smaller than the value
obtained with no restriction on the space between fibers. It lies slightly
below the flow stress of the hexagonal array subjected to  tension at $0^0$ 
($\Sigma_0=87.9 $ MPa). However it lies above the flow stress of the square
array under tension at $0^0$ or $45^0$ ($\Sigma_0=79.6$ MPa) and of the 
hexagonal array under tension at $45^0$.
\item[2.]
When the matrix  is elastic-plastic with linear hardening, the effective
hardening modulus drops significantly ~: $H = 9382$ MPa
(estimation error = $123.8$ MPa), instead of $10 002$ MPa. But it is still
much higher than the hardening modulus predicted with the hexagonal
 array case ($H=7100 $ MPa at $0^0$, $H=7420 $ MPa at $45^0$).
\end{itemize}
In conclusion, it seems that the ``safety coating" around the fibers leads to
a decrease in the overall mechanical properties of the composite, at least
at the volume fraction which has been investigated.

\subsubsection{Influence of the shape of the fibers}

The above analyses show that the overall flow stress of the composite and, 
to a lesser extent,
its overall hardening depend primarily on the tortuosity of shear bands
passing through the matrix. Obviously, the volume fraction of the reinforcing 
phase plays a role in the possibility that such bands are formed, but for
a fixed volume fraction, significant differences arise from the differences
in the patterning of bands. These shear bands are locked or deviated by 
the fibers. The overall flow stress of the composite can 
(empirically) be related to the length of the shortest path passing through the
matrix and making an angle of approximately $45^0$ with the tensile direction. It
can be expected that the {\it shape} of the fibers, which act as  
"shear bands barriers", is important in their capacity to inhibit shear
bands. The shape of fibers is important at two levels. First it affects the
arrangement of fibers in the unit cell. For instance, it can be favorable
to clustering of particles, leaving large areas of inclusions-free matrix
where plastic strain is likely to localize. At a smaller scale 
an elongated particle perpendicular to a shear band will form an effective
barrier.
\vskip 0.3cm
Random microstructures were generated with three shapes of fibers~:
circular, elliptical (aspect ratio= 3.333), equilateral triangles. The volume 
fraction was $47.5 \% $. The unit cells contained 64 fibers and were
discretized into  $1024 \times 1024$ pixels.
The center of the fibers and their orientation were chosen randomly, subject to
the contraints of periodicity, impenetrability and given volume fraction. A
minimal space of four pixels between two fibers was imposed to correctly
capture the high strain gradients in the matrix between two neighboring
reinforcements. For each  fiber shape, 10 different configurations were 
tested. The results of the numerical simulations are given in Table 9.
\vskip 0.3cm
The Young's modulus is not significantly affected by the shape of the inclusions, at least
for this particular volume fraction and for the contrast of elastic
properties which was investigated (investigation of the percolation threshold 
for highly contrasted phases would probably lead to different conclusions).
The mean flow stress of the composite with elliptical inclusions is close to
that of the composite with circular inclusions ($ 0.9\%$ higher). However 
the flow stress is significantly
higher for the composite with triangular inclusions 
($5.2\%$ higher). This "hardening" effect can be attributed to the fact that
at a given volume fraction triangles form more efficient barriers to shear
band formation. This efficiency can be related to the  length of the
projection of the fiber orthogonally to the shear bands. 
The minimal length, the maximal length, and the
average length over all possible orientations are reported in Table 10
for each shape of fibers at a given area $s$. For circular
fibers
these three quantities are equal  to the radius of the fiber (2 $\sqrt{s / \pi}$).

\subsection{Penetrable fibers}

When the matrix is elastic ideally plastic, 
the overall response of the composite is strongly influenced by the 
existence of continuous paths in the matrix, connected from one cell to the
other. The contiguity of the matrix obviously plays a crucial role in the
formation of these paths, which are ruled out when the matrix 
is not contiguous.
\vskip 0.3cm
In order to study this effect, different configurations at different volume
fractions were generated with {\it penetrable} fibers. The centers of the
fibers were first chosen at random. Then the 
volume fraction of the reinforcing phase was controlled by increasing the 
radius of the fibers (all fibers at a given volume fraction had identical radius).
The matrix was assumed to be elastic perfectly plastic. 
The results of the simulations can be analyzed as follows~:
\begin{itemize}
\item[1.]
When the fiber volume fraction is small, shear bands can be passed through
the matrix. According to Drucker's remark, the resulting overall flow stress 
of the composite coincides with the flow  stress of the 
matrix under plane strains,  $2 \sigma_0 / \sqrt{3}$. However,
when the fiber volume fraction is very small, a nearly homogeneous
deformation of the matrix is more favorable (less energy is dissipated in
the 
plastic deformation) and no strain concentration is observed. Then the
overall flow stress of the composite stands between the flow stress of the
matrix $\sigma_0$ and the flow stress of the matrix under plane strains 
($2 \sigma_0 / \sqrt{3}$).

\item[2.]
Over a certain radius, straight shear bands cannot be passed through the
matrix. 
For a
given geometrical distribution of fibers, this radius is half of the maximal distance
between adjacent parallel lines passing through the centers of the fibers 
and inclined at $\pm 45^o$ on the tensile direction. 
Periodic continuous paths can again be passed through the matrix 
but they are tortuous. The bands where the plastic strain concentrates 
have a nonvanishing width. 
The stress-strain response
of the composite again reaches a limit value, one
higher than the flow stress of the matrix under plane strains. The overall
flow stress
increases with the volume fraction of fibers, and the
increase is closely related to the tortuosity of the "shear"
bands.
\item[3.]
When the fibers percolate and form a contiguous phase, the matrix loses 
contiguity. 
No periodic continuous path can be passed through the matrix. 
This leads to a drastic modification
of the stress-strain curve of the composite, which is no longer limited. 
The composite behaves asymptotically as an elastic plastic material with linear hardening
.
\end{itemize}
\subsection{Complex microstructures}
To illustrate the capability of the method to deal with complex
microstructures, we have considered a real microstructure taken from the
work of {  Bornert} \cite{BOR96} (see also \cite{BOR94}). The materials studied 
in \cite{BOR94} were two-phase iron/silver blends, manufactured with powder metallurgy 
techniques. The digital image was obtained by Scanning Electron Microscopy. 
The microstructure is shown in Figure \ref{bornert} (a).
Clearly meshing this microstructure for application of the FEM would be a considerable task. 
The present numerical method can handle such a microstructure 
as easily as the simpler ones shown in previous examples. 
In the numerical simulation each phase is considered
elastic-plastic following a $J_2$-flow theory with isotropic hardening of
the Von-Mises type. The stress/strain curves for each constituent under uniaxial
tension are shown in Figure \ref{bornert} (c). The applied loading is
uniaxial tension in the horizontal direction. The map of equivalent strain 
is shown in Figure \ref{bornert} (b) at an overall strain $E_{11} = 3.3 \%$. 
In the soft phase (silver in white) the strain is organized in bands which cannot develop over 
long distances due to the presence of the hard phase (iron in black).   
A full comparison between simulated and experimental strain maps is difficult
to perform essentially  because the
numerical calculations are two-dimensional whereas the real material is
three-dimensional in nature. Only the surface of the specimen is observed and it
is in a state of plane stress, whereas the calculations are performed 
assuming a state of generalized plane strains. In addition the material below the
 surface plays a significant role on the deformation of the surface
itself. The variations between the arrangement of the phases at the surface and 
below the surface is not taken into account by the numerical model.    
\section{Concluding remarks}
A new numerical technique has been developed to investigate the local and
overall response of nonlinear composites. The advantages of the method are
the following~:
\begin{itemize}
\item[1.] Images of microstructures can be directly used in the analysis,
which avoids meshing the microstructure. Complex microstructures can be
investigated. Part of the efficiency of the method is due to the use of FFT
packages.
\item[2.] The iterative procedure does not require the formation or
inversion of a stiffness matrix.
\item[3.] Convergence is fast.
\end{itemize}
However the method has some limitations.
\begin{itemize}
\item[1.] Convergence is not ensured for materials containing voids or rigid
inclusions.
\item[2.] The number of degrees of freedom is high by comparison with the FEM
(typically an image with $1024 \times 1024$ pixels is required to deal with 64 
fibers). The method can be implemented only on  computers
with high memory capabilities.
\end{itemize} 

\noindent{\bf Acknowledgements}. Most computations were carried out at the
Institut M\'editerran\'een de Technologie in Marseille; the funds being
provided by the PACA region. The other computations were carried out at the 
Institut du D\'eveloppement et des Ressources en Informatique Scientifique
funded by CNRS. The authors are indebted to Michel Bornert for fruitful
discussions and for providing the image of the microstructure shown in 
\ref{bornert} (a).

\newpage
\appendix
\section{Green's operator of a linear elastic material
}
\label{appendix1}
The auxiliary problem of a homogeneous material with stiffness ${\bf c}^0$ 
subject to a periodic polarization field $\taubf$ plays an important role 
in the method which has been proposed. Its solution, which can be found in several
textbooks ({\it e.g.} {\sc Mura} \cite{MUR87}), can be expressed in 
terms of the Fourier transform of the polarization field by means of the 
Fourier transform of the Green's operator of the following systems of equations
\begin{equation}
\left.
\begin{array}{c}
 \sigmabf({\bf x}) =
{\bf c}^0 : {\bf \varepsilonbf} ( {\bf u}^*({\bf x}) ) + \taubf({\bf x})
\quad \forall {\bf x} \in V \\ \\
{\rm div} \sigmabf({\bf x}) = {\bf 0} \quad \forall {\bf x} \in V ,\quad \sigmabf{\bf .n} \ -\# ,\quad \ {\bf u}^\ast \#
\end{array}
\right\}
\label{ap1}
\end{equation}
In Fourier space, these equations take the form
\begin{equation}
\hat \sigma_{ij}(\xibf) = {\rm i}\
c^0_{ijkh}\ \xi_ h\ \hat {u}^\ast_ {k} (\xibf) + \hat {\tau}_{ij}(
\xibf),\quad
 {\rm i}\ \hat {\sigma}_{ij}( \xibf)\ \xi_ j =
0 .
\label{ap2}
\end{equation}
(It is hoped that the index $i$ will not be confused with the complex number 
${\rm i}=\sqrt{-1}$). Eliminating $ \hat {\sigma}_{ ij} $ between the two 
equations in (\ref{ap2}) yields
$$ K^0_{ik} (\xibf). u^*_k = \hat {\tau}_{ ij}
(\xibf)\ \xi_j,
$$
where $ {\bf K}^0(\xibf)$ denotes the acoustic tensor  of the homogeneous material, $ K^0_{ik} (\xibf)
= c^0_{ijkh}\ \xi_h\ \xi_j . $ 
Then
$$ \hat u^\ast_ k(\xibf) = {\rm i}\ N^0_{ki} (\xibf)\ \hat {\tau}_{ ij}
(\xibf)\ \xi_j  = {{\rm i} \over 2}\ (N^0_{ki} (\xibf)\ \xi_j  + N^0_{kj} (\xibf)\ \xi_i )\ \hat \tau_{ij} (\xibf) , $$
where the symmetry of $\taubf$ has been used and where 
$ {\bf N}^0 (\xibf)$ denotes the inverse of $ {\bf K}^0(\xibf)$. Therefore
\beg
\hat{\varepsilon}_{kh}(u^\ast) =  
{{\rm i} \over 2} \left( \xi_
h\ \hat {u}^\ast_ k(\xibf) + \xi_ k\ \hat {u}^\ast_ h(\xibf) \right)
=  \hat{\Gamma}^0_{khij} (\xibf)\ \hat{ \tau}_{ij}(\xibf),  
\label{3.3}
\en
with
\beg
\hat{\Gamma}^0_{khij}  =  
{1 \over 4} \left( N^0_{hi}(\xibf)\ \xi_j\ \xi_k + 
N^0_{ki}(\xibf)\ \xi_j\ \xi_h  + N^0_{hj}(\xibf)\ \xi_i\ \xi_k + 
N^0_{kj} (\xibf)\ \xi_i\ \xi_h \right) , 
\label{3.4}
\en
and
\beg
 \hat{\tau}_{ij}(\xibf) = <\tau_{ij}({\bf x}) e^{- {\rm i}
\xibf.{\bf x}}>. 
\label{3.5}
\en
The strain field induced at each point $\bf x$ of the unit cell $V$ by an initial stress $\taubf$  
can be determined from (\ref{3.3}), (\ref{3.4}) and (\ref{3.5}). These formulas give the explicit form of the operator 
${\bf \Gamma}^{0}$ and of the operation $*$ considered in section 2:
$$ \epsilonbf({\bf u} ^ \ast) = - {\bf \Gamma}^0 * \taubf . $$
Deatailed expressions of ${\bf \Gamma}^0$ can be found in {\sc Mura}
\cite{MUR87} for different types of anisotropy for the reference medium. 
Its expression is particularly simple when the material is isotropic with Lam\'e coefficients 
$\lambda_0$ and $\mu_0$; the above expression becomes :
$$ c^0_{ijkh} = \lambda^0 \delta_{ij} \delta_{kh} + \mu^0 \left(\delta_{ik} \delta_{jh} + \delta_{ih} \delta_{jk} \right). 
$$
$$K^0_{ij} ({\bf \xibf}) =  \left( \lambda^0+ \mu^ 0 \right) \xi_i \xi_j + 
\mu^0 \vert {\bf \xibf} \vert^ 2 \delta_{ij}
$$
$$N^0_{ij} ({\bf \xibf}) =  {1 \over \mu^ 0 \vert {\bf \xibf} \vert^ 2} \left( \delta_{ij} - {\xi_i\ \xi_j \over \vert {\bf \xibf} \vert^2} {\lambda^0 +
\mu^0 \over \lambda^0 + 2\mu^0} \right). $$
Therefore:
$$
\hat{\Gamma}_{khij} ({\bf \xibf}) =  {1 \over 4 \mu^ 0 \vert {\bf \xibf} 
\vert^ 2} \left( \delta_{ki} \xi_ h \xi_ j +
\delta_{hi} \xi_ k \xi_ j\ + \delta_{kj} \xi_ h \xi_ i + \delta_{ hj} \xi_ k 
\xi_ i
\right) -  {\lambda^0+ \mu^0 \over
\mu^0(\lambda^0+ 2\mu^0)} \ {\xi_i \xi_j \xi_k \xi_h \over \vert {\bf \xibf} \vert^4} 
$$

\newpage
\section{Imposing a macroscopic stress  direction.}
\label{appendix3}
In the above described algorithm the overall strain is
prescribed by assessing the value of the Fourier transform of the strain
field at the zero frequency~: 
$$ \hat{\varepsilonbf}({\bf 0}) = {\bf E}.$$
It is often convenient (or  necessary) to impose the overall stress
$\bfS$, 
rather than the overall strain $\bfE$. A typical example is provided by 
uniaxial tension
in the transverse direction as described by (\ref{ut}). In strongly 
nonlinear problems it is
even necessary to impose only the direction of the overall stress  and to drive
the loading by means of an auxiliary parameter (arc length method). 
The algorithm can be modified to account for loadings in the form 
\beg
 {\bf \Sigma} \ = \ k \  {\bf S}_0 \ \ \hbox{\rm and} \ 
\ {\bf E}:{\bf S}_0 = t,
\label{a31}
\en
where ${\bf S}_0$ is the prescribed direction of overall stress (by direction of
stress we refer to a direction in the 6-dimensional space of stresses), $k$ is
the unknown level of overall stress  and $t$, which serves as a
loading parameter, is the component of the overall strain in 
this direction. Then, the overall strain and stress  
${\bf E}^{i}$ et ${\bf \Sigma}^{i}$ have to be determined by means of 
(\ref{a31}).
For this purpose, at iterate $i$, $\sigmabf^{i-1}$ and $\varepsilonbf^{i-1}$ being known, 
the loading level $t^i$ being known but $k^i$ being unknown,  
${\bf E}^{i}$ and ${\bf \Sigma}^{i}$ are subject to~:
\beg
\left.
\begin{array}{c}
{\bf \Sigma}^{i} - {\bf c}^0:{\bf E}^{i} = 
<\sigmabf^{i-1}>-{\bf c}^0:<\varepsilonbf^{i-1}> \\ \\
{\bf \Sigma}^{i} = k^{i} {\bf S}_0,\quad 
{\bf E}^{i}:{\bf S}_0 = t^{i}
\end{array}
\right\}
\label{a32}
\en
Elimination of ${\bf \Sigma}^{i}$ yields
\beg
{\bf E}^{i} = 
k^{i} {{\bf c}^0}^{-1}:{\bf S}_0 \ - \ {{\bf c}^0}^{-1}:<\sigmabf^{i-1}>
\ + \ <\varepsilonbf^{i-1}>
\label{a33}
\en
and 
$$
k^{i} = 
{
t^{i} + ({{\bf c}^0}^{-1}:<\sigmabf^{i-1}>-<\varepsilonbf^{i-1}>):{\bf S}_0
\over
{{\bf c}^0}^{-1}:{\bf S}_0:{\bf S}_0 
} 
$$
Therefore the modification brought into the algorithm (\ref{alg4})
is an additional step to determine ${\bf E}^{i}$ according to (\ref{a33}), 
which is then prescribed as the
overall strain through~:
$$ {\hat \varepsilonbf^{i}}({\bf 0}) = {\bf E}^{i}.$$

It is worth noting that the
condition 
$<\varepsilonbf^i> = {\bf E}^i$ is met at each step of the iterative procedure,
 whereas the equality
$<\sigmabf^i> = \bfS^i$
is met only at convergence. The difference arises from the fact that  
$\sigmabf^i$ is deduced from the
constitutive law, whereas $ \bfS^i$ is deduced from (\ref{a32}). 
Indeed, once convergence is reached, one has 
$${\bf E}^{i} = {\bf E}^{i-1} = <\varepsilonbf^{i-1}>,$$
and, according to (\ref{a32}), 
$ {\bf \Sigma}^{i} = <{\bfs}^{i-1}> = <{\bfs}^{i}>$ .

\newpage

\section{Radial return algorithm}
\label{appendix2}
The equations  governing a plastic material obeying a $J_2$ flow theory
with isotropic hardening read~:
\begin{equation}
{\dot \sigmabf} = {\bf c}:({\dot \varepsilonbf}-{\dot \varepsilonbf}^{p}),
\quad
\dot {\varepsilonbf}^{p} = 
	{3 \over 2} \ {\dot p} \ { \displaystyle {\bf s} \over  \displaystyle \sigma_{eq} },
\label{B0}
\en
\beg
\left.
\begin{array}{c}
\dot p = 0 \ {\rm when} \ {\sigma_{eq}}-{\sigma_{0}}(p) < 0,\\
\\
 \dot p > 0 \ {\rm when} \ {\sigma_{eq}}-{\sigma_{0}}(p) = 0 .
\end{array}[B
\right\}
\label{B1}
\end{equation}
$\varepsilonbf^{p}$ is the plastic strain, 
$p$ is the equivalent plastic strain $ \dot p = \left( {2 \over 3} \ 
	   { \dot {\varepsilonbf}^{p} } : 
	   { \dot {\varepsilonbf}^{p} } \right)^{1/2}$.
$\bf c$ is the stiffness tensor, assumed to be isotropic and  characterized 
by a bulk modulus $k$ and a shear modulus $\mu$.
\vskip 0.3cm
Time is discretized into intervals $[t^n,t^{n+1}]$.
$F^n$ denotes the value of a function $F$ at time $t^n$. 
$\varepsilonbf^n$, $\sigmabf^n$ and $p^n$ denote the strain, stress and
equivalent plastic strain at time $t^n$.
Given the mechanical fields at step $n$, and given the strain field 
$\varepsilonbf^{n+1}$ at step $n+1$, the constitutive law amounts 
to finding the stress field $\sigmabf^{n+1}$ and the equivalent plastic 
strain field $p^{n+1}$.
Replacing time differentiation by a finite difference in
(\ref{B0}) 
provides 
$$
\sigmabf^{n+1} - \sigmabf^{n} = 
{\bf c} \ : \ \left(
\varepsilonbf^{n+1} - \varepsilonbf^{n} 
- {\dot{\varepsilonbf}^p}{ }^{n+1} \times (t^{n+1}-t^n) 
\right) .
$$
The elastic prediction is
\beg
{\sigmabf}_T^{n+1} \ = \ \sigmabf^n + {\bf c}:
\left( \varepsilonbf^{n+1}-\varepsilonbf^{n} \right).
\label{B2}
\en
After due account of plastic incompressibility, (\ref{B2}) gives
$$
\sigmabf^{n+1} = {\sigmabf}_T^{n+1} 
- 2  (t^{n+1}-t^n)\  \mu \ {\dot{\varepsilonbf}^p}{ }^{n+1}. 
$$
Alternatively, making use of the flow rule (\ref{B0}) and of the 
decomposition of $\sigmabf^{n+1}$ into a spherical stress and deviator
stress
\begin{equation}
 {\rm tr}(\sigmabf^{n+1}) = {\rm tr}({\sigmabf}_T^{n+1})
 = {\rm tr}(\sigmabf^{n}) + 3k \ {\rm tr}(\varepsilonbf^{n+1}-\varepsilonbf^{n}) 
\label{B3}
\en 
\beg
{\bf s}^{n+1} = {\bf s}_T^{n+1}  - {3  (t^{n+1}-t^n) \ \mu \
\dot{p}^{n+1} \ 
 \over \displaystyle \sigma^{n+1}_{eq} } \ {\bf s}^{n+1}
\label{B4}
\end{equation}
(\ref{B3}) can be re-written, assuming that there are no initial stresses 
or strains  at time $t^0$, 
$${\rm tr}(\sigmabf^{n+1}) = 3k \ {\rm tr}(\varepsilonbf^{n+1})$$
The radial return method is based on the observation that, according to
(\ref{B4}), the deviators  
${\bf s}^{n+1}$ and ${\bf s}_T^{n+1} $ are proportional.
The Von Mises stresses  associated with ${\sigmabf^{n+1}}$ 
and $\sigmabf_T^{n+1} $ are therefore related through~:
\beg
\sigma^{n+1}_{eq} \ \ = \ \ (\sigma_T^{n+1} )_{eq} \ 
- \ 3 \mu \ (p^{n+1}-p^n) 
\label{B5}
\en
\begin{itemize}
\item[-] 
If $(\sigma_T^{n+1} )_{eq} < \sigma_0(p^n)$,  the step is purely
elastic,
$$\sigmabf^{n+1} = \sigmabf_T^{n+1},\quad p^{n+1}= p^n.$$
\item[-]  
If $(\sigma_T^{n+1} )_{eq} \geq \sigma_0(p^n)$, 
the material plastifies at step $n+1$, $\sigma^{n+1}_{eq}=
\sigma_0(p^{n+1})$  and (\ref{B5}) reduces to~:
$$ \sigma_0(p^{n+1}) + 3 \mu \ p^{n+1} = (\sigma_T^{n+1} )_{eq} + 
3 \mu \ p^n 
$$
Assuming that  hardening is positive (no softening), the function 
$h(p) = \sigma_0(p) + 3 \mu  p$ can be inverted to give
\beg
p^{n+1} = h^{-1}\big( (\sigma_T^{n+1} )_{eq} + 3 \mu p^n \big)
\label{B6}
\en
The case of linear hardening leads to simple inversion. Indeed, in this
case, $\sigma_0(p) = \sigma_0 + H  p$ and (\ref{B6}) reduces to 
$$ p^{n+1} = {3 \mu \over H + 3 \mu } \ p^n + 
{ (\sigma_T^{n+1} )_{eq}-\sigma_0 
 \over  H + 3 \mu }. $$
The case of a perfectly plastic material, corresponding to $H=0$, is
covered by the above relation.
When $h^{-1}$ is not available in a closed form, it can be approximated 
by linear interpolation. When  $k \in \ [h(p_l),h(p_{l+1})]$, 
$p=h^{-1}(k)$ is approximated by 
$p_l + (k-h(p_l)) {\displaystyle  p_{l+1}-p_l \over \displaystyle 
h(p_{l+1})-h(p_l)} $.
\end{itemize}
Finally, the algorithm used in our computations reads~:
\beg
\left.
\begin{array}{ll}
& \varepsilonbf^n, \ \sigmabf^n, \ p^n, \ \varepsilonbf^{n+1}  \ 
{\rm being \ known}, \\ \\
{\rm Compute} &  {\bf s}_T^{n+1} = 
{\bf s}^n + 2 \mu (\varepsilonbf^{n+1}-\varepsilonbf^n) ,\\
&  (\sigma_T^{n+1} )_{eq} = 
\left({3 \over 2} {\bf s}_T^{n+1} : {\bf s}_T^{n+1} \right)^{1/2} 
\\ \\
{\rm Test} & {If} \quad (\sigma_T^{n+1} )_{eq} < \sigma_0(p^n) \\
 & \qquad
\begin{array}{rl}
	& p^{n+1} = p^n \\ 
	& {\bf s}^{n+1} = {\bf s}_T^{n+1}
\end{array}
\\
 & {Else} \\
 & \qquad 
\begin{array}{rl}
	& p^{n+1} = h^{-1}\big( ({\sigmabf}_T^{n+1} )_{eq} + 
		3 \mu p^n \big) \\ 
	& {\bf s}^{n+1} = 
		 \ 
		{\displaystyle \sigma_0(p^{n+1}) \over 
		\displaystyle (\sigma_T^{n+1})_{eq} }\  {\bf s}_T^{n+1}
\end{array}
\\
{\rm End \ of \ test} &\\ \\
{\rm Update} & {\rm tr}(\sigmabf^{n+1}) = {\rm tr}(\sigmabf^{n}) + 
3k \ {\rm tr}(\varepsilonbf^{n+1} - \varepsilonbf^{n})  \\ 
& \sigmabf^{n+1} = {1 \over 3}{\rm tr}(\sigmabf^{n+1}) \ \bkI {\rm d} \ \  + \ \ {\bf s}^{n+1}
\end{array}
\right\}
\end{equation}

\clearpage
%

\clearpage

%
\clearpage
\begin{figure}
  \parbox{15cm}{
    \parbox{7cm}{
      \includegraphics[width=7cm]{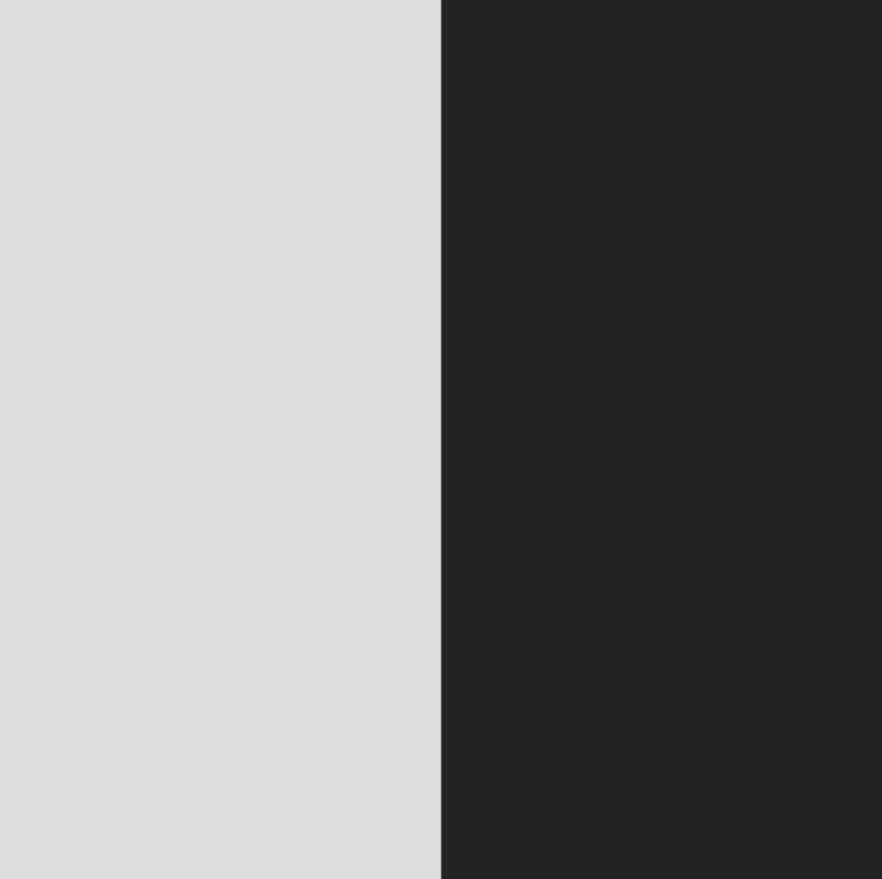}
      \begin{center}
        {\it (a)} 
      \end{center}
    }
    \parbox{8.5cm}{
      \includegraphics[width=8.5cm]{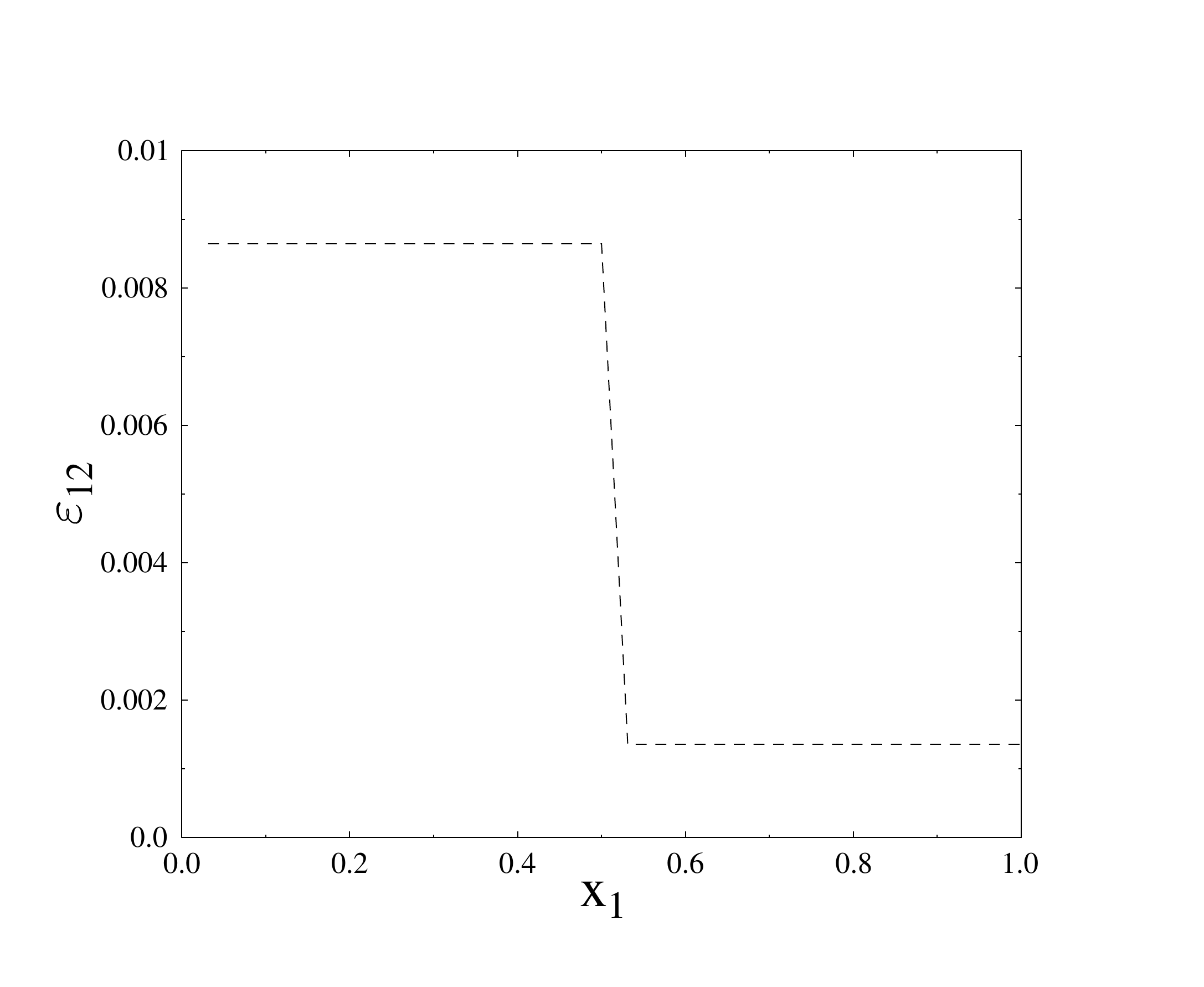}
      \begin{center}
        {\it (b)} 
      \end{center}
    }
  }
  \caption{Two-phase laminate. 
    ($E^1 \ = \ 68.9 \ {\rm GPa},\quad \nu^1=0.35$,
    $E^2 \ = \ 400 \ {\rm GPa},\quad \nu^2 \ = \ 0.23$ ). Volume fraction
    of both phases $50 \%$.
    Spatial resolution $32 \times 32$ pixels. Applied loading: pure shear in the
    plane $(x_1,x_2)$. (a): map of the local strain field $\varepsilon_{12}$. 
    (b) : cut through $\varepsilon_{12}$ 
    along an arbitrary horizontal line.}
  \label{layer}
\end{figure}

%
\clearpage
\begin{figure} 
  \begin{center}
    \parbox[t]{6cm} {
      \includegraphics[width=6cm]{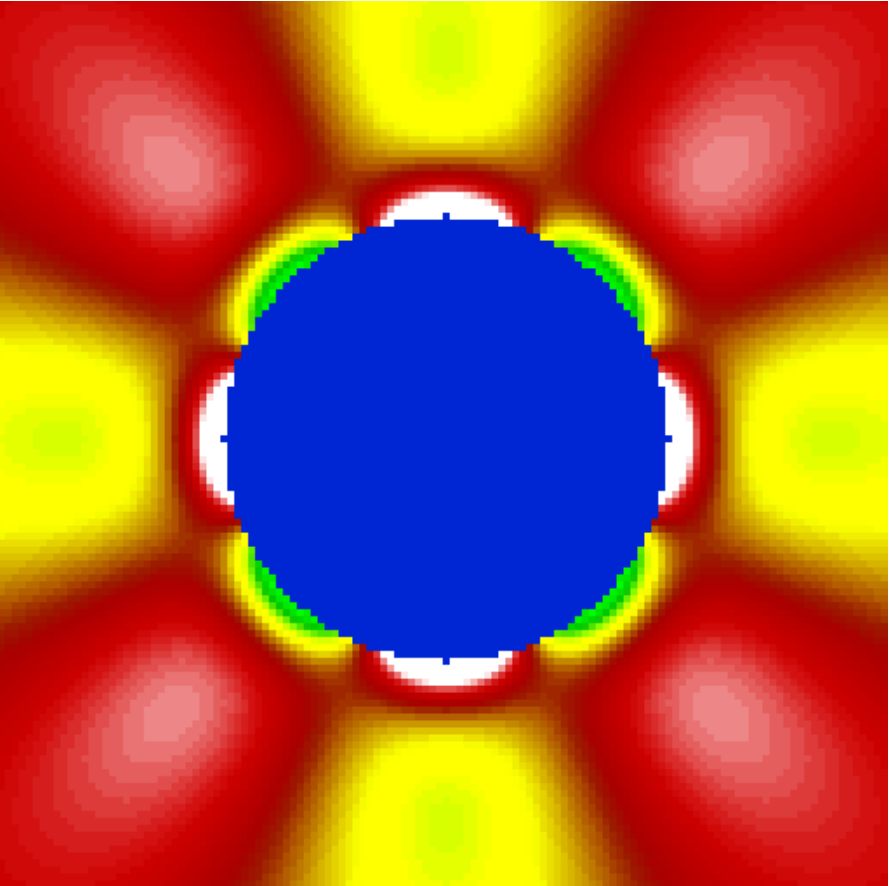}
      \begin{center}
        {\it (a)} 
      \end{center}
    }
    \hskip 0.5cm
    \parbox[t]{6cm} {
      \includegraphics[width=6cm]{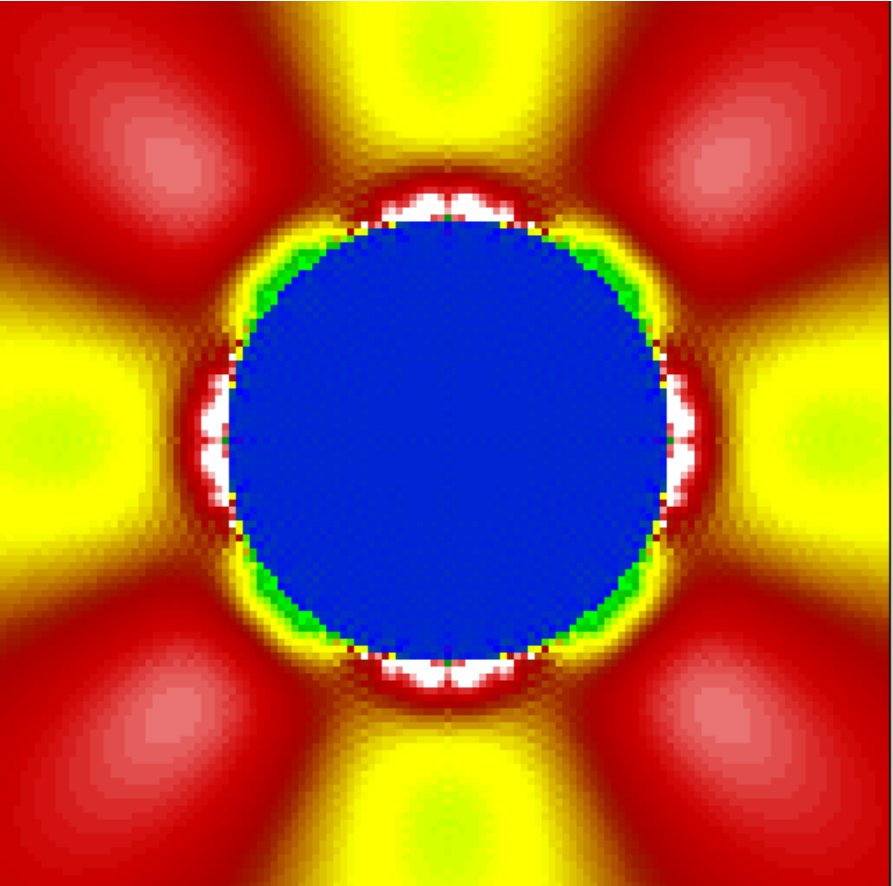}
      \begin{center}
        {\it (b)} 
      \end{center}
    }
    \hskip 0.2cm
    \parbox[t]{1.cm} {
      \includegraphics[width=1.2cm]{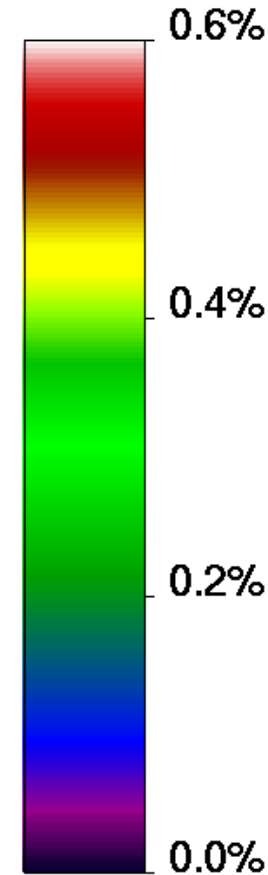}
      \begin{center}
        {\it ~ } 
      \end{center}
    }
  \end{center}
  \begin{center}
   \parbox{14cm} {
      \includegraphics[width=14cm]{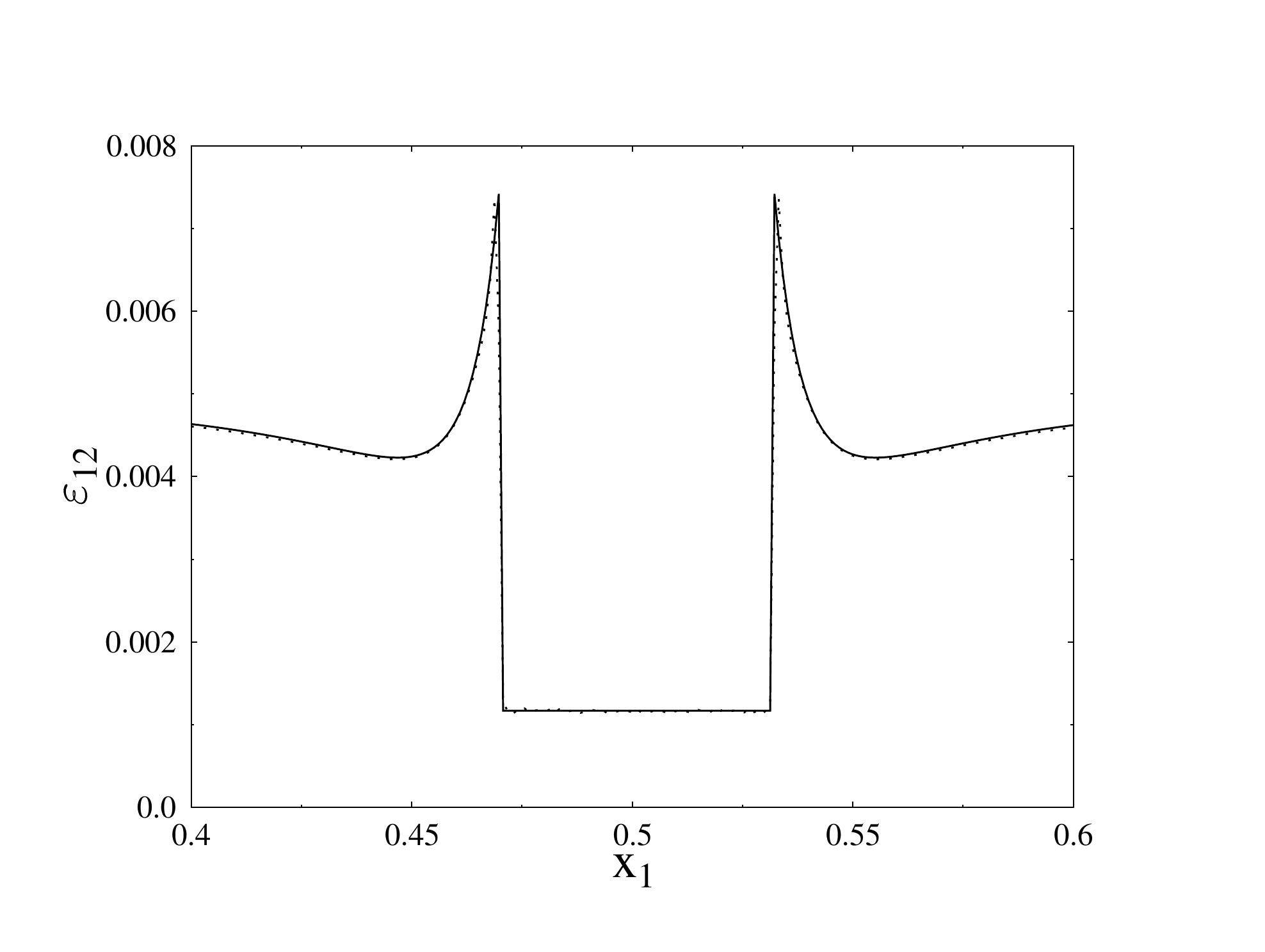}
      \begin{center}
        {\it (c)} 
      \end{center}
   }
  \end{center}
\caption{Circular fiber in a matrix with elastic mismatch 
between the phases
($E^m \ = \ 68.9 \ {\rm GPa},\quad \nu^m=0.35$ 
and
$E^f \ = \ 400 \ {\rm GPa},\quad \nu^f \ = \ 0.23$ ).
Shear loading: $E_{12} = 0.5 \%, \ E_{ij}=0 \ \forall \ (i,j) \ne (1,2)$.
Maps of  the local strain field $\varepsilon_{12}$. 
(a): analytical solution, (b): numerical simulation. 
Spatial resolution $1024 \times 1024$ pixels.
(c): cut through $\varepsilon_{12}$ at $x_2=0$.
Dotted line: analytical solution, solid line:
numerical simulation.
}
\label{infini}

\end{figure}


\clearpage
\begin{figure}
  \begin{center}
    \parbox{15.5cm}{
      \parbox{7.5cm} {
        \includegraphics[width=7.5cm]{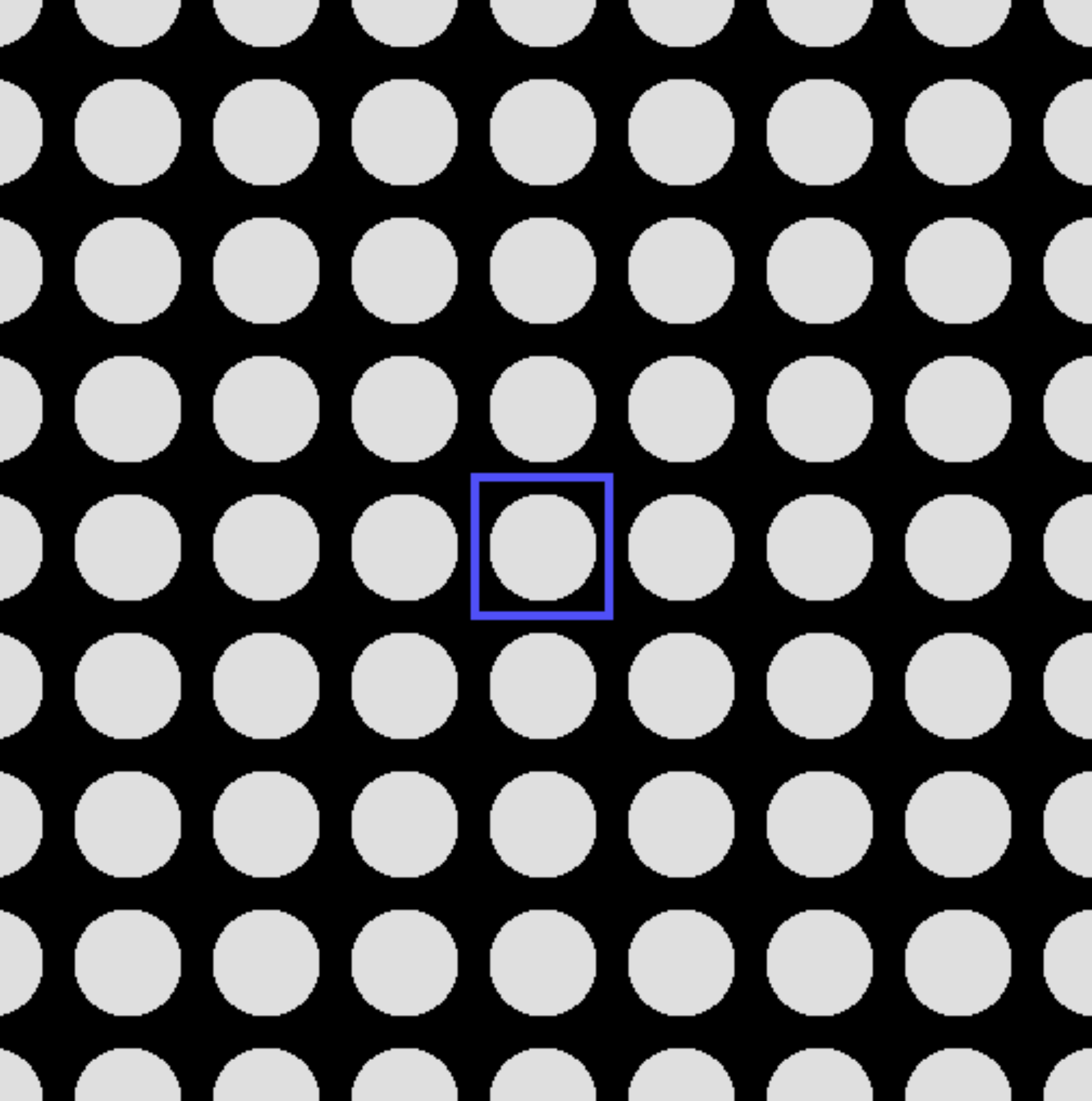}
        \begin{center}
          {\it (a)} 
        \end{center}
      }
      \hspace{0.3cm}
      \parbox{7.5cm} {
        \includegraphics[width=7.5cm]{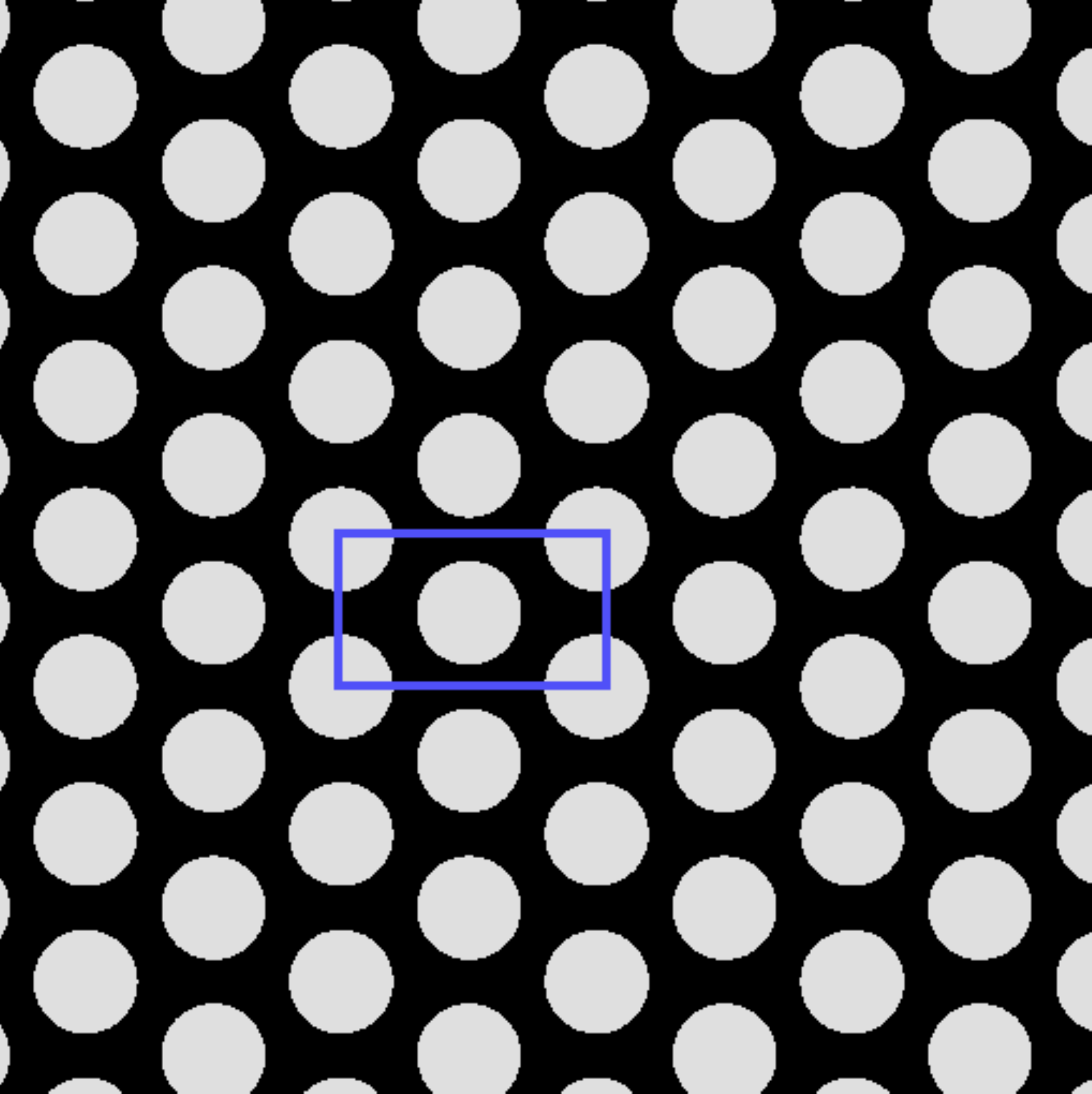}
        \begin{center}
          {\it (b) }
        \end{center}
      }
    }
  \end{center} 
  \caption{{Standard fiber distributions. (a) : square lattice, the unit cell 
      contains one fiber. (b): hexagonal lattice,  the unit cell contains 
      $1+ 4\times {1 \over 4} = 2$ fibers.}}
  \label{std}
\end{figure}

\begin{figure} 
\begin{center}{
\leavevmode
\parbox{7.5cm} {
\includegraphics[width=7.5cm]{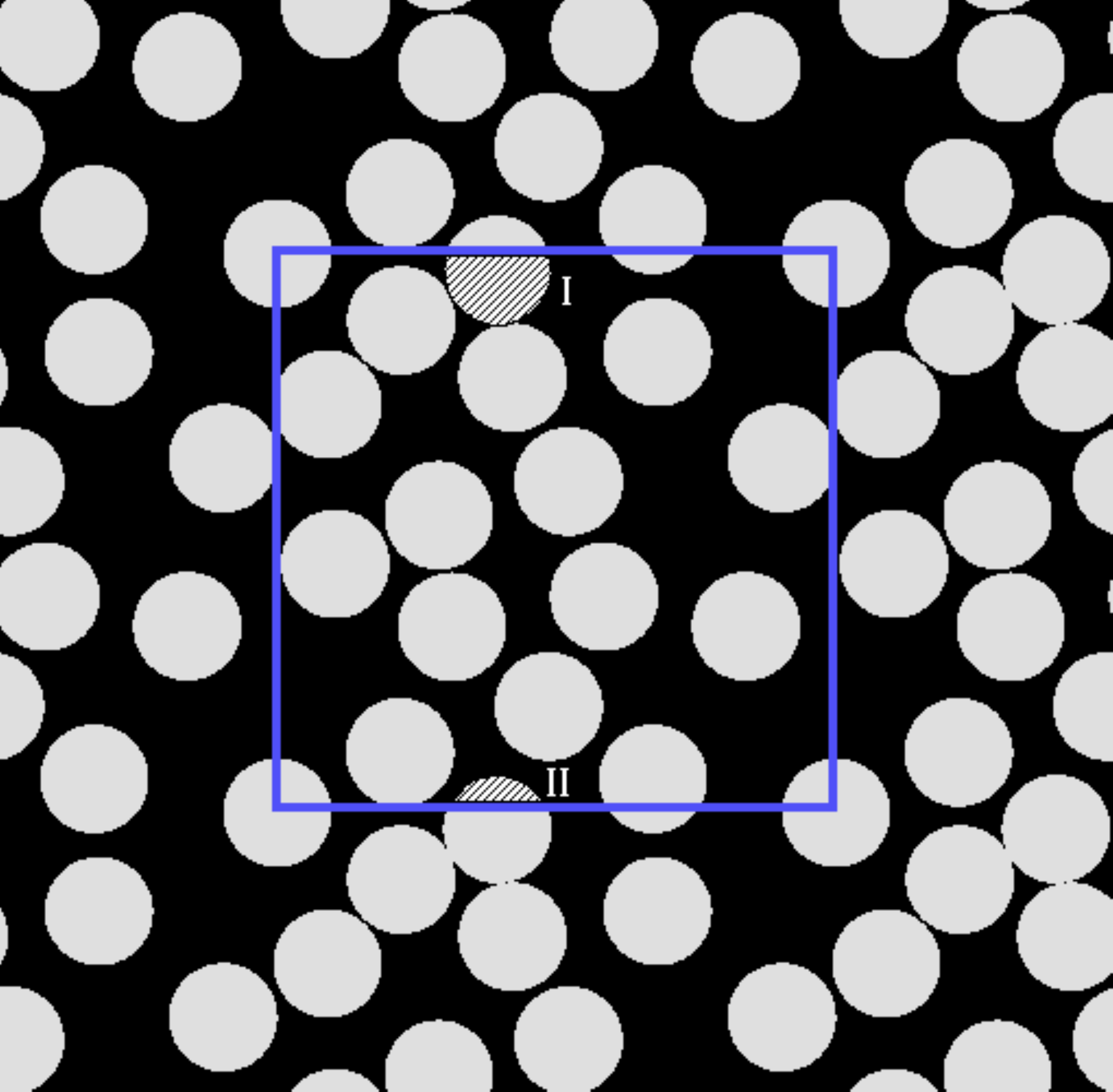}
}
\caption{{Periodic unit cell containing 16 circular fibers randomly placed.}}
\label{fcar}
} \end{center} \end{figure}

\clearpage
\begin{figure} 
  \begin{center} 
    \parbox{16cm} {
      \begin{center} 
        \parbox{12cm}{
          \includegraphics[width=12cm]{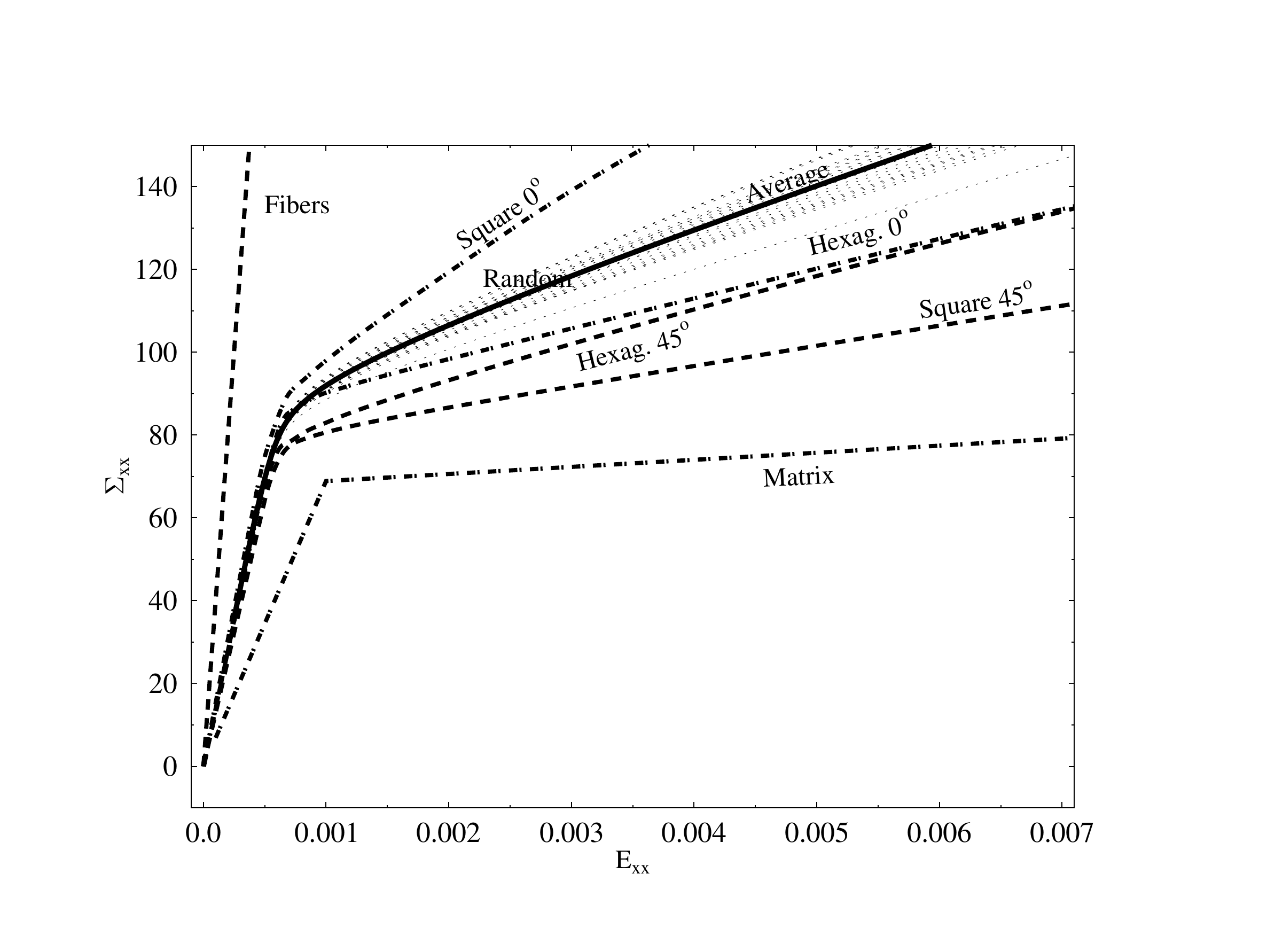}
          \begin{center}
            {\it (a) }
          \end{center}
        }
      \end{center}
      \begin{center} 
        \parbox{12cm}{
          \includegraphics[width=12cm]{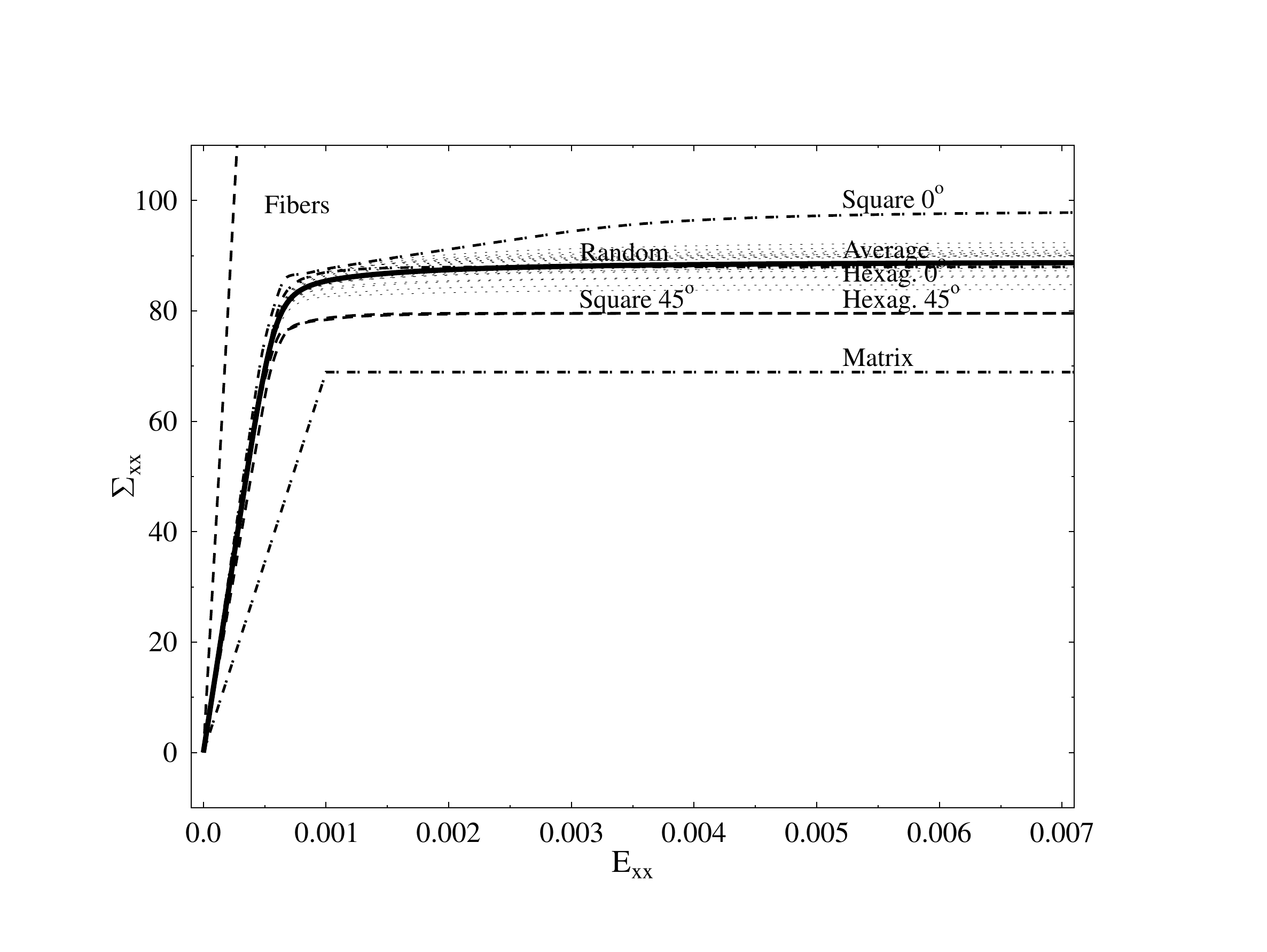}
          \begin{center}
            { \it (b) }
          \end{center}
        }
      \end{center}
    }
  \end{center}
  \caption{{Overall stress-strain response computed with the present method. 
Volume fraction of fibers: $47.5\%$. 
(a) Matrix with linear hardening. (b) Ideally plastic matrix
Dotted lines: 23 configurations of 64 identical circular fibers 
placed randomly in the r.v.e. Thick solid line: average of the responses of
the random configurations. 
Square $0^0$ (resp: Square $45^0$): fibers placed at the nodes of a square lattice,
 tension at $0^0$ (resp. $45^0$). 
Hexag. $0^0$ (resp: Hexag. $45^0$): fibers placed at the nodes of a hexagonal lattice, 
tension at $0^0$ (resp. $45^0$). 
}}
  \label{courbes}
\end{figure}

\clearpage
\begin{figure} 
\begin{center} 
{
\leavevmode
\parbox[t]{6cm} {
\includegraphics[width=6cm]{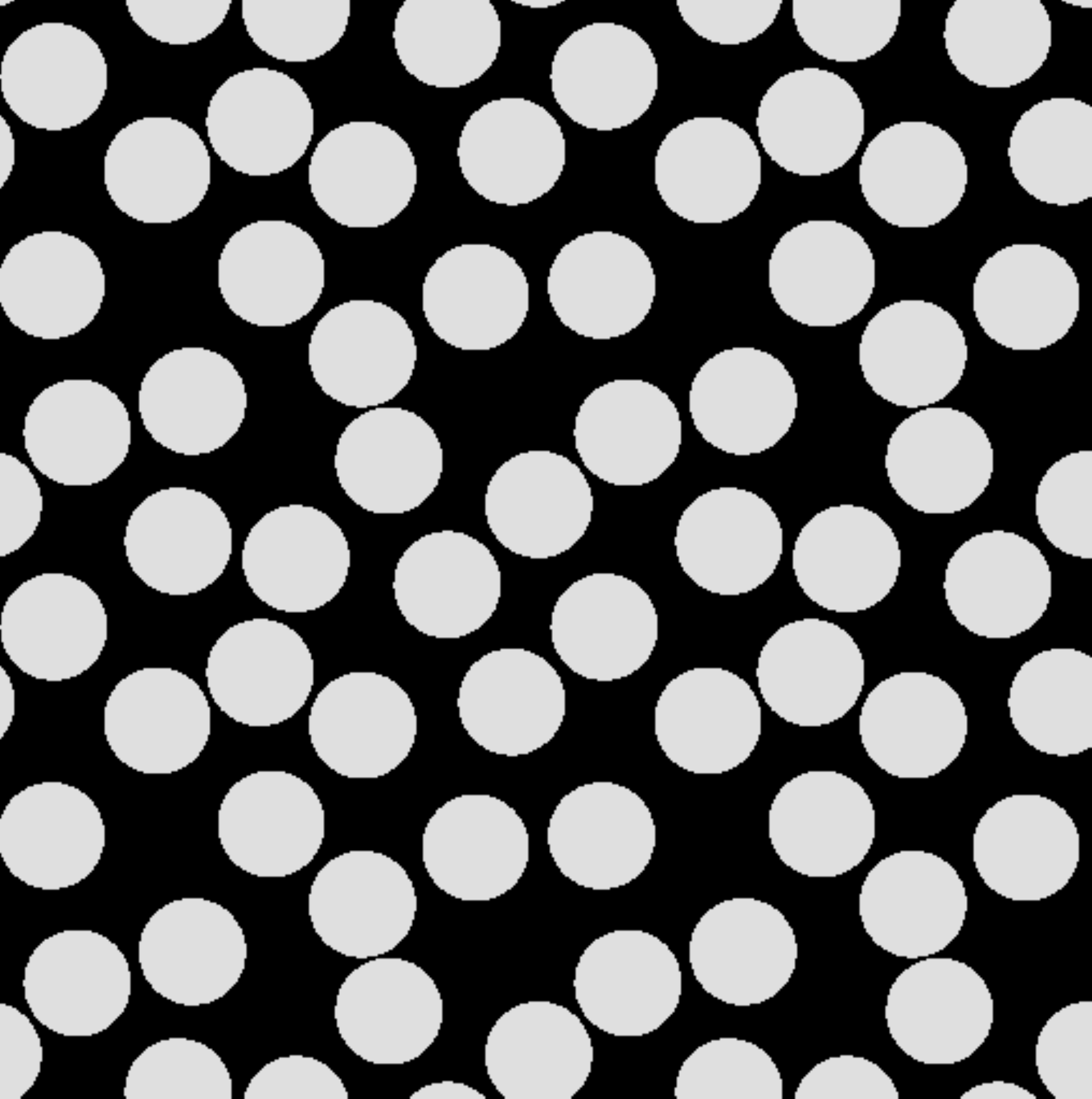}
\begin{center}
(a)
\end{center}
}
\hskip 0.5cm
\parbox[t]{6cm} {
\includegraphics[width=6cm]{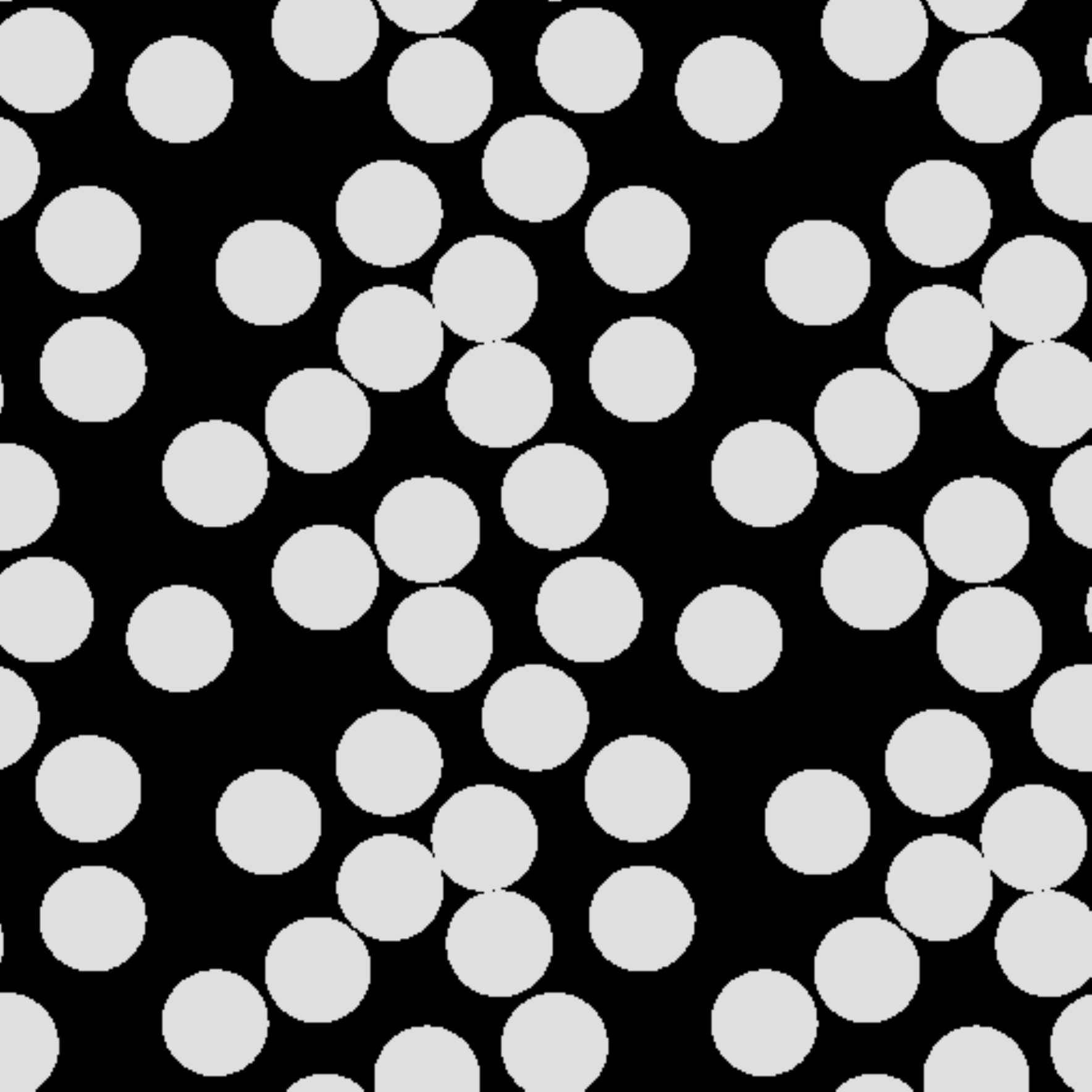}
\begin{center}
(d)
\end{center}
}
\hskip 0.2cm
\parbox[t]{1cm} {\hglue 1cm}
\vskip 0.1cm
\parbox[t]{6cm} {
\includegraphics[width=6cm]{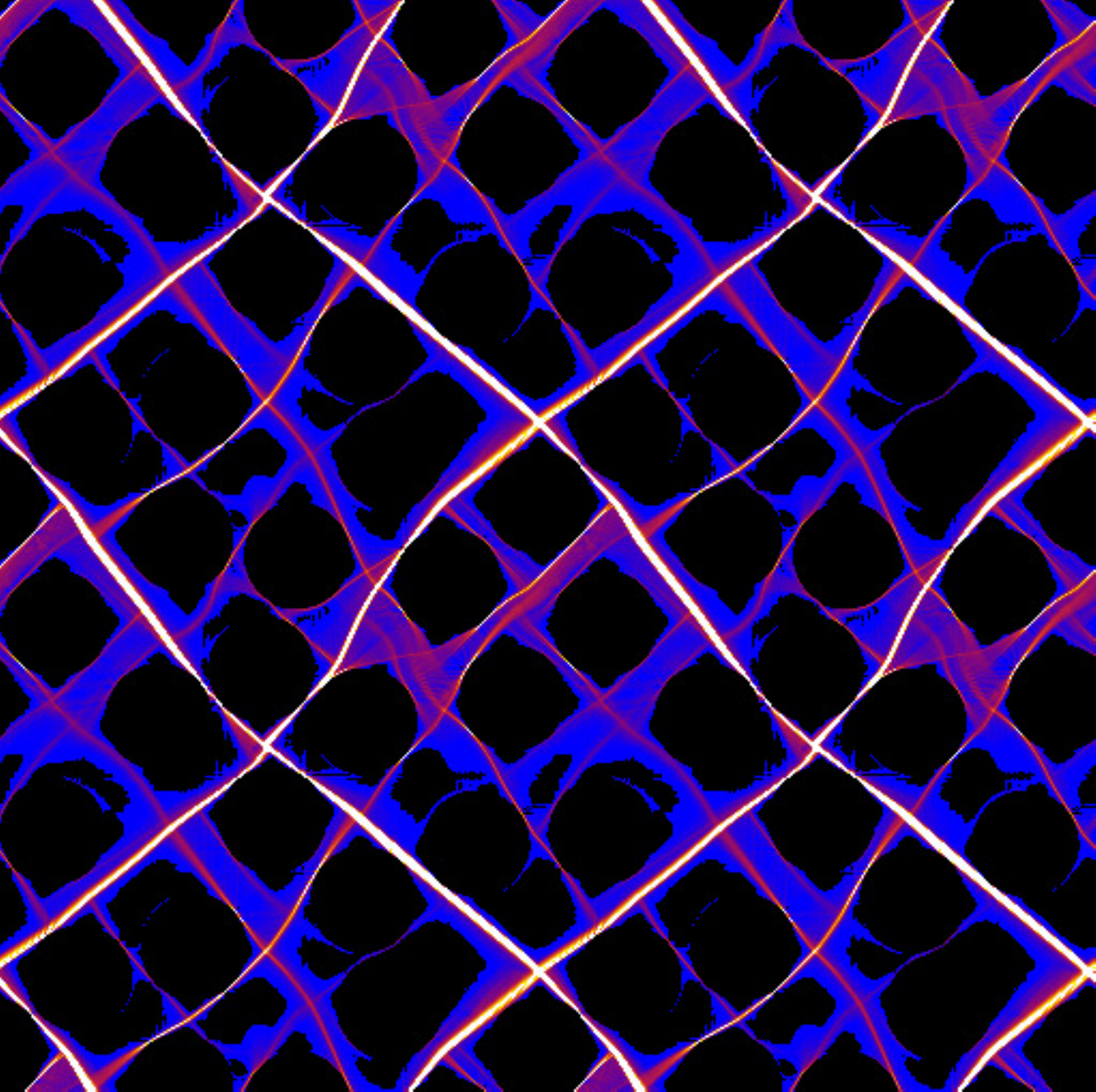}
\begin{center}
(b)
\end{center}}
\hskip 0.5cm
\parbox[t]{6cm} {
\includegraphics[width=6cm]{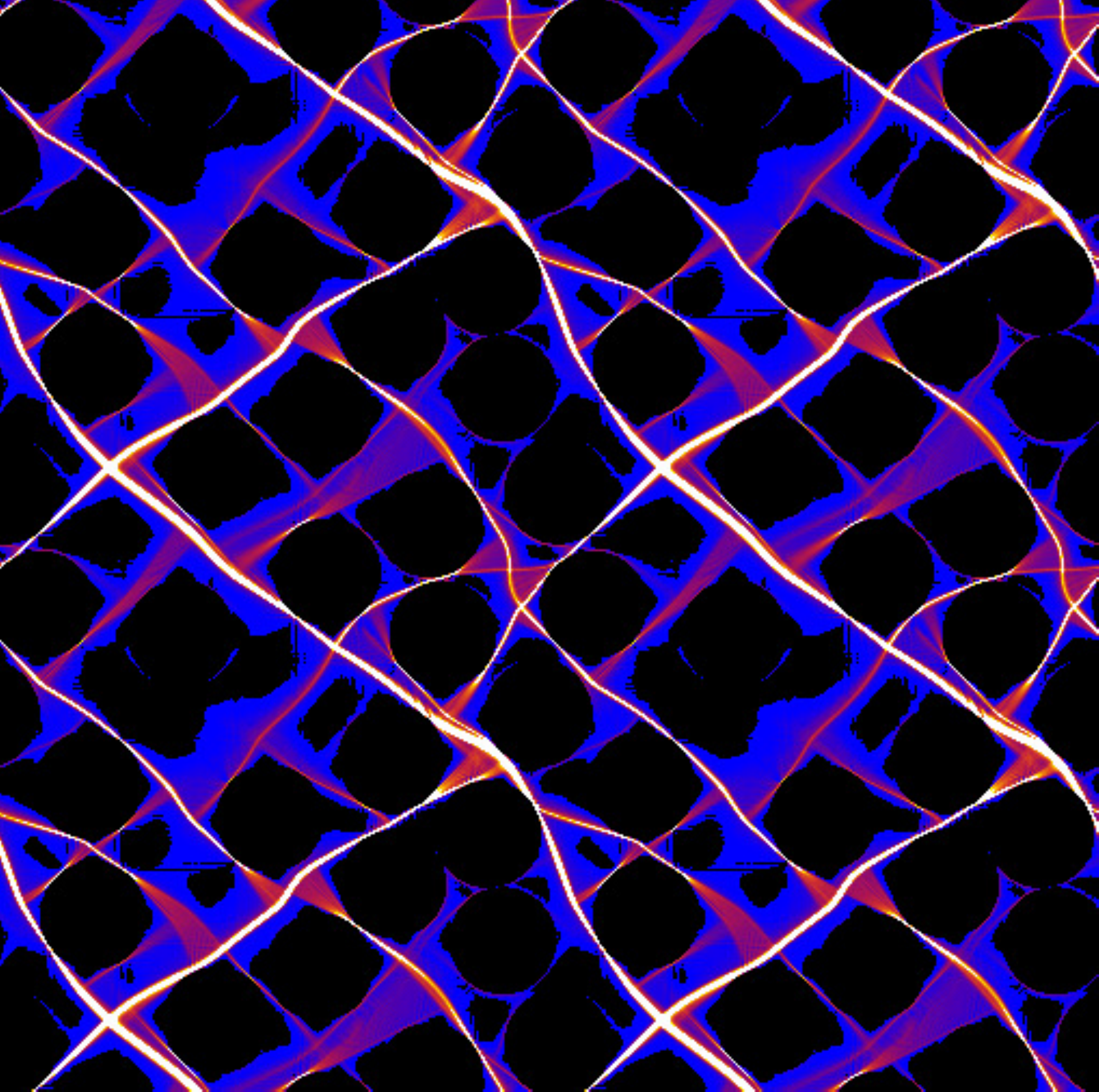}
\begin{center}
(e)
\end{center}
}
\hskip 0.2cm
\parbox[t]{1cm} {
\includegraphics[width=1.2cm]{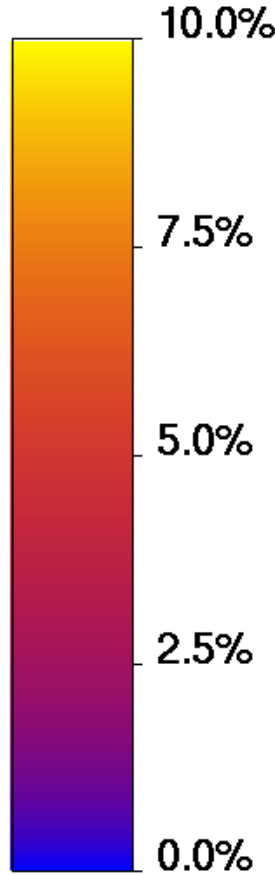}}
\vskip 0.1cm
\parbox[t]{6cm} {
\includegraphics[width=6cm]{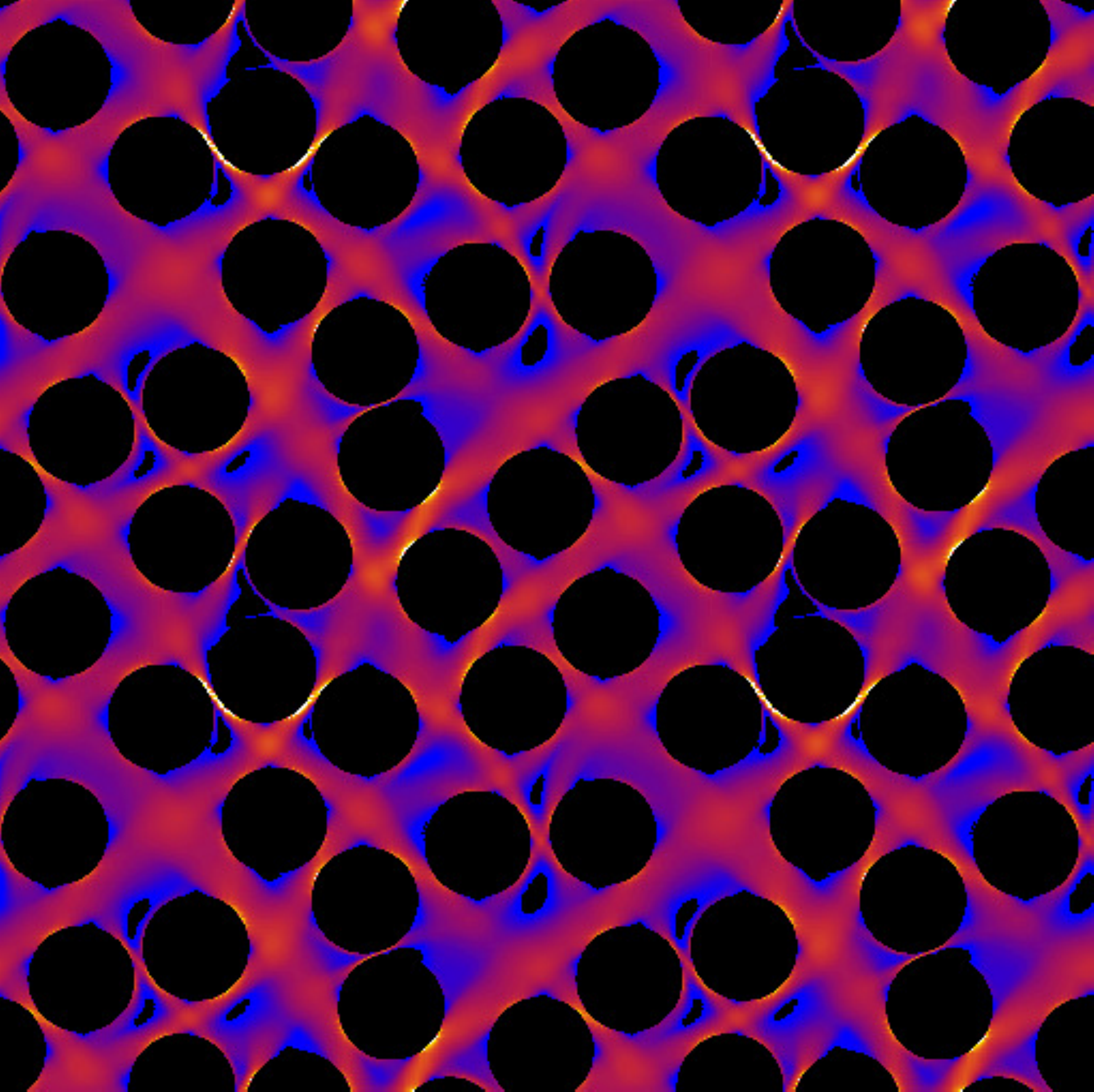}
\begin{center}
(c)
\end{center}}
\hskip 0.5cm
\parbox[t]{6cm} {
\includegraphics[width=6cm]{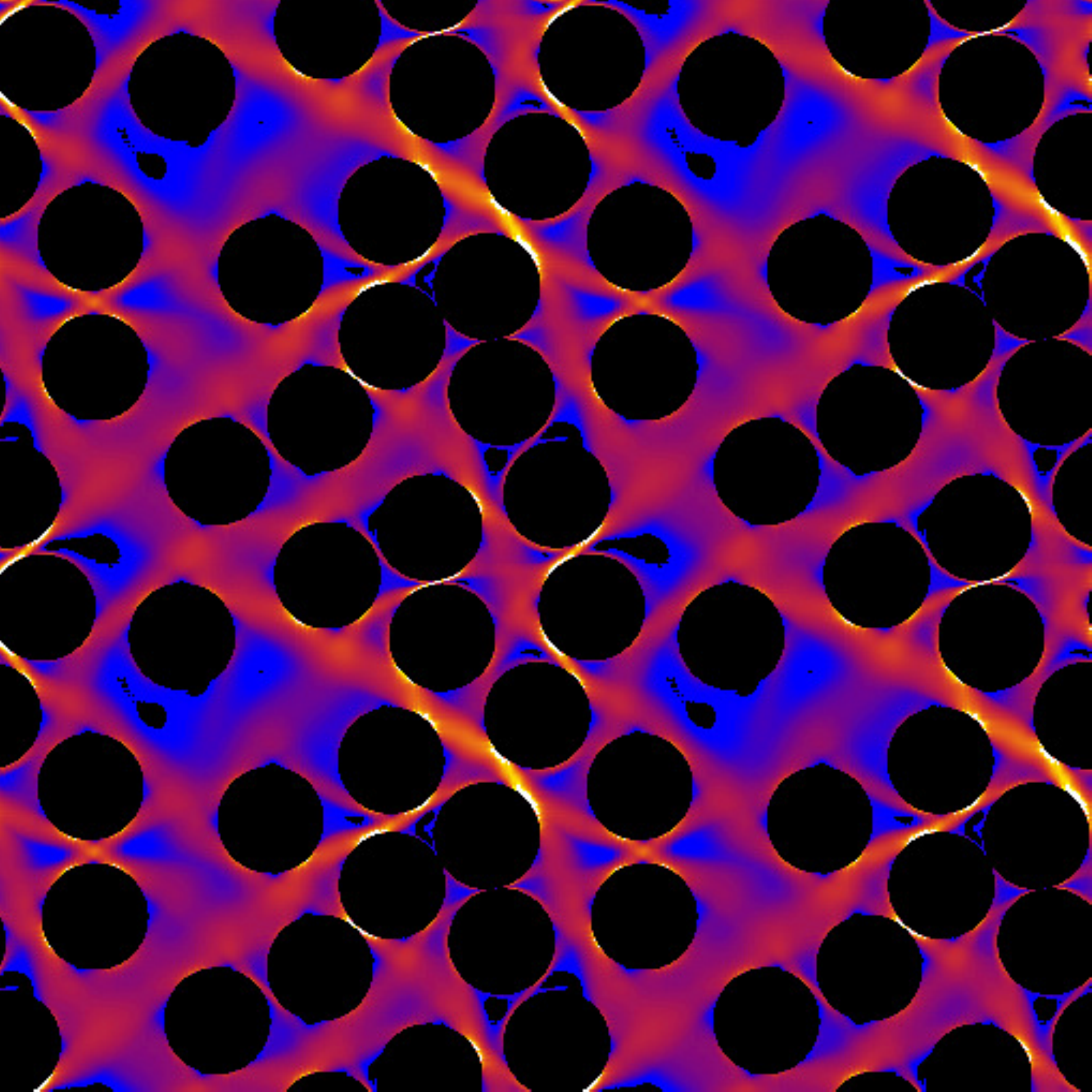}
\begin{center}
(f)
\end{center}
}
\hskip 0.2cm
\parbox[t]{1cm} {
\includegraphics[width=1.2cm]{lut2-eps-converted-to.pdf}}
}
\end{center}
\caption{{Two different microstructures ( (a) and (d) )  
and the corresponding
plastic strain maps.  The matrix is elastic - ideally plastic in (b) and
(f). The matrix is elastic plastic with linear hardening  in (c) and (f). 
Transverse uniaxial tension.
Overall strain $E_{11}= 1 \%$. $0\%$ strains are 
displayed in black, $10\%$ strains 
(and more) are displayed in white.
Straight slip bands can form easily in configuration (a). 
The slip bands are more tortuous in  
configuration (d). When the matrix is ideally plastic the overall flow stress 
of configuration (d) is $6.4\%$ higher than the flow stress in configuration
(a).
}}
\label{pstrain}
\end{figure}

\clearpage
\begin{figure}
  \begin{center}
    \parbox{12cm} {
      \includegraphics[width=12cm]{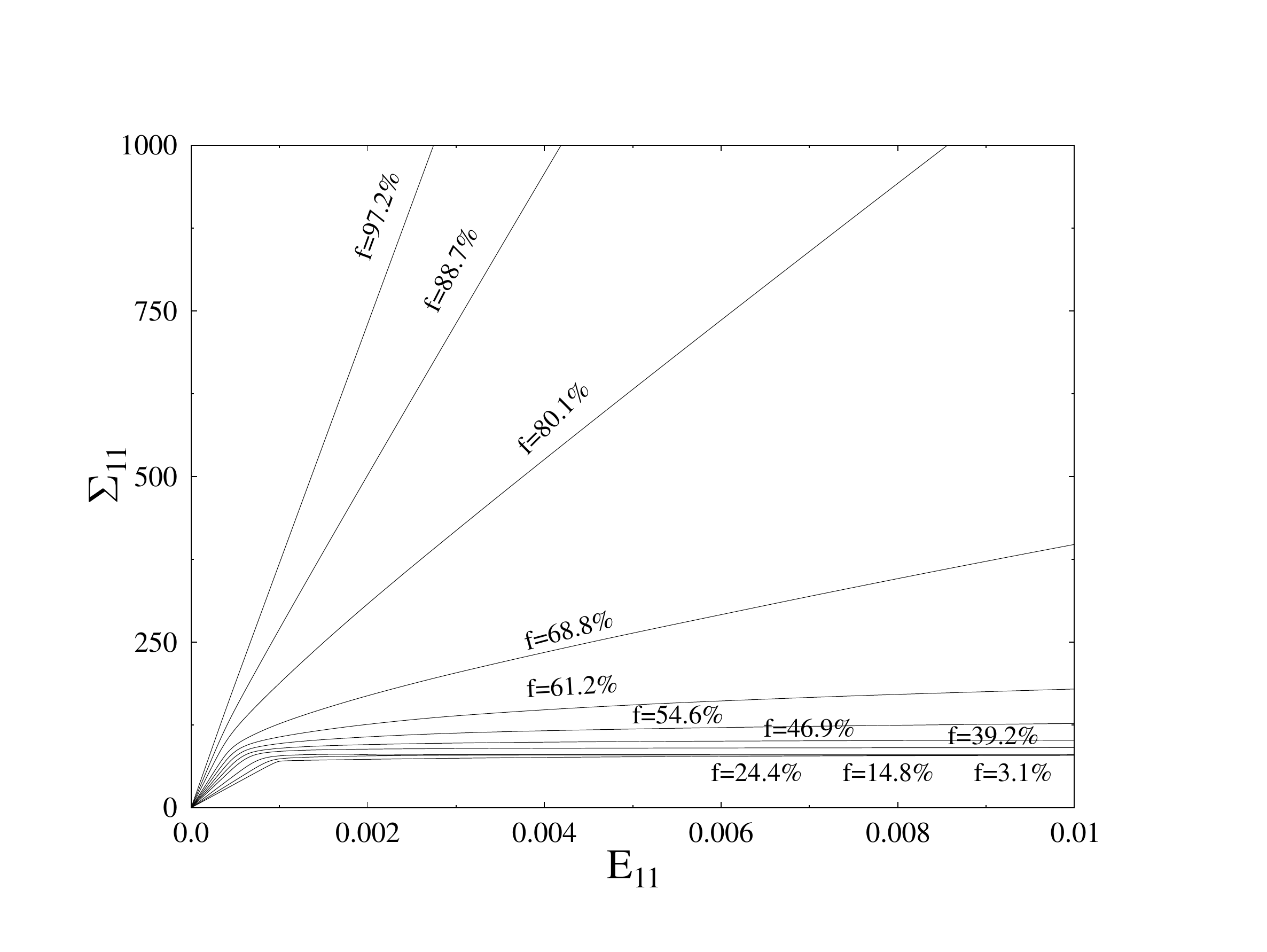}
    }
  \end{center}
\caption{{
Overall stress-strain response of fiber reinforced materials at different
fiber volume fraction $f$. 100 penetrable circular fibers with increasing 
radius in an elastic ideally plastic matrix.
($E^m \ = \ 68.9 \ {\rm GPa},\quad \nu^m=0.35 
\quad \sigma_0 \ = \ 68.9 \ MPa $ 
and
$E^f \ = \ 400 \ {\rm GPa},\quad \nu^f \ = \ 0.23$ ).
Tension at $0^0$.
}}
\label{figmt}
\end{figure}

%
\clearpage
\begin{figure}
  \parbox{17cm}{
    \parbox[t]{7.2cm}{
      \includegraphics[width=7.2cm]{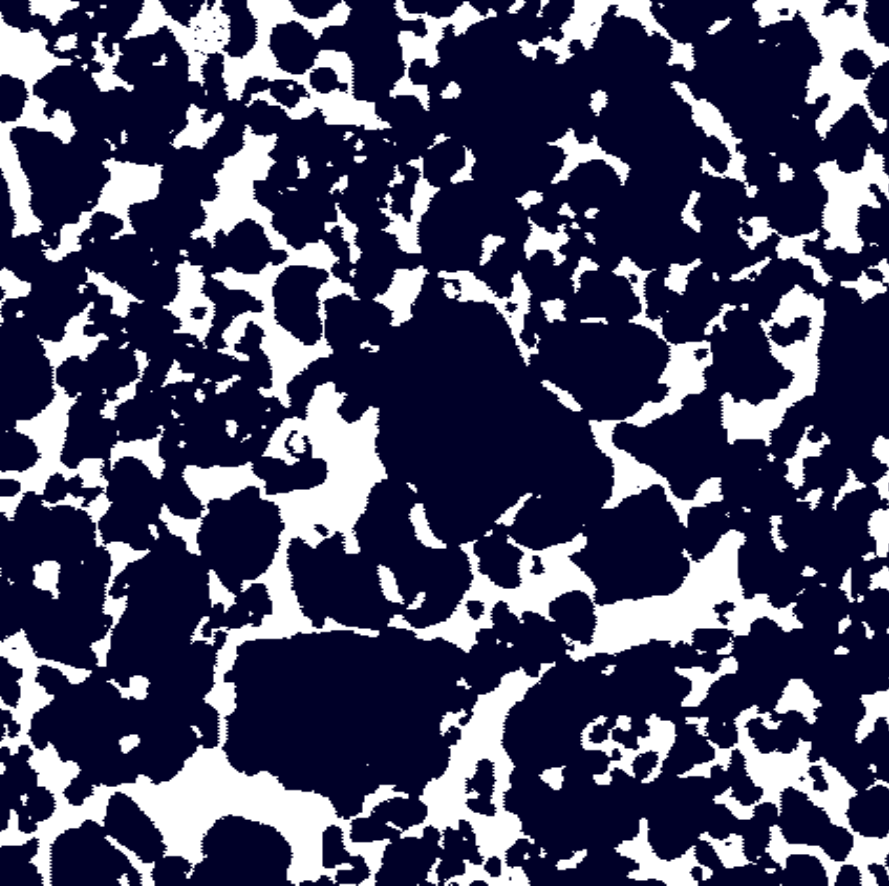}
      \vskip 0.5cm
      \begin{center}
        {\it (a)} 
      \end{center}
    }
    \hspace{0.3cm}
    \parbox[t]{7.2cm} {
      \includegraphics[width=7.2cm]{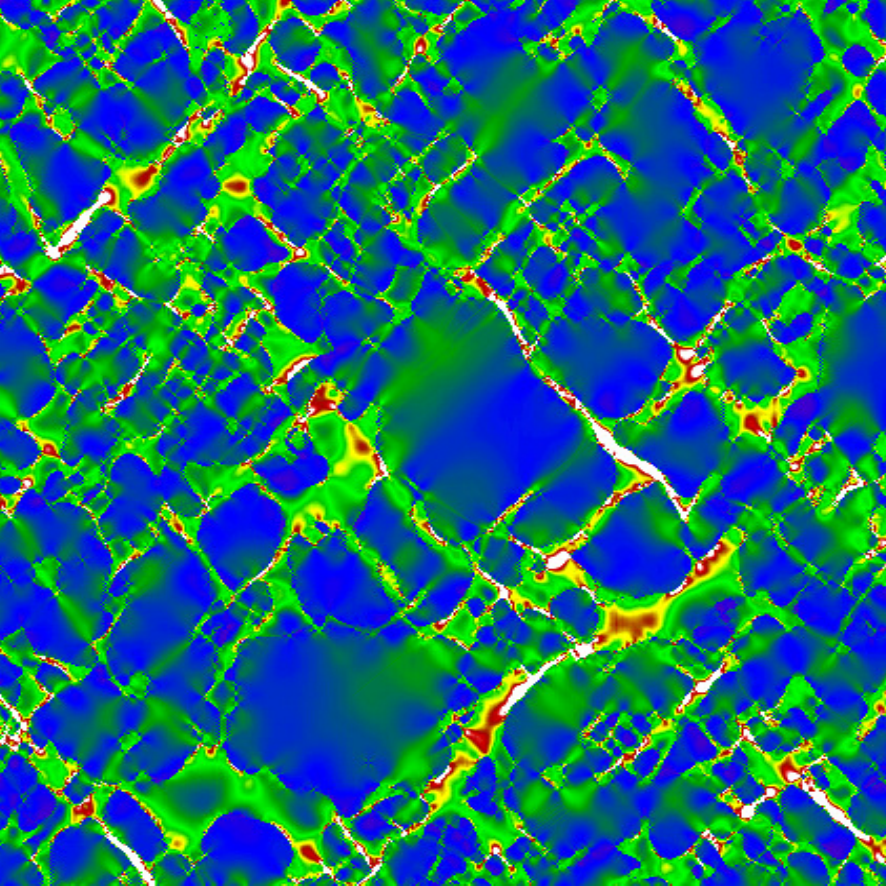}
      \vskip 0.5cm
      \begin{center}
        {\it (b)} 
      \end{center}
    }
    \hspace{0.1cm}
    \parbox[t]{1.8cm}{
      \includegraphics[width=1.2cm]{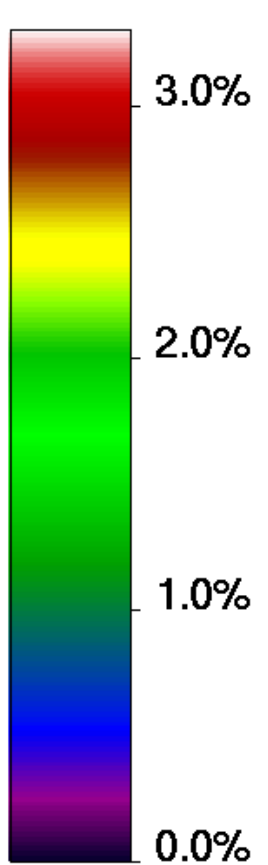}
    }
    \vskip 1cm
    \parbox{12cm} {
      \includegraphics[width=12cm]{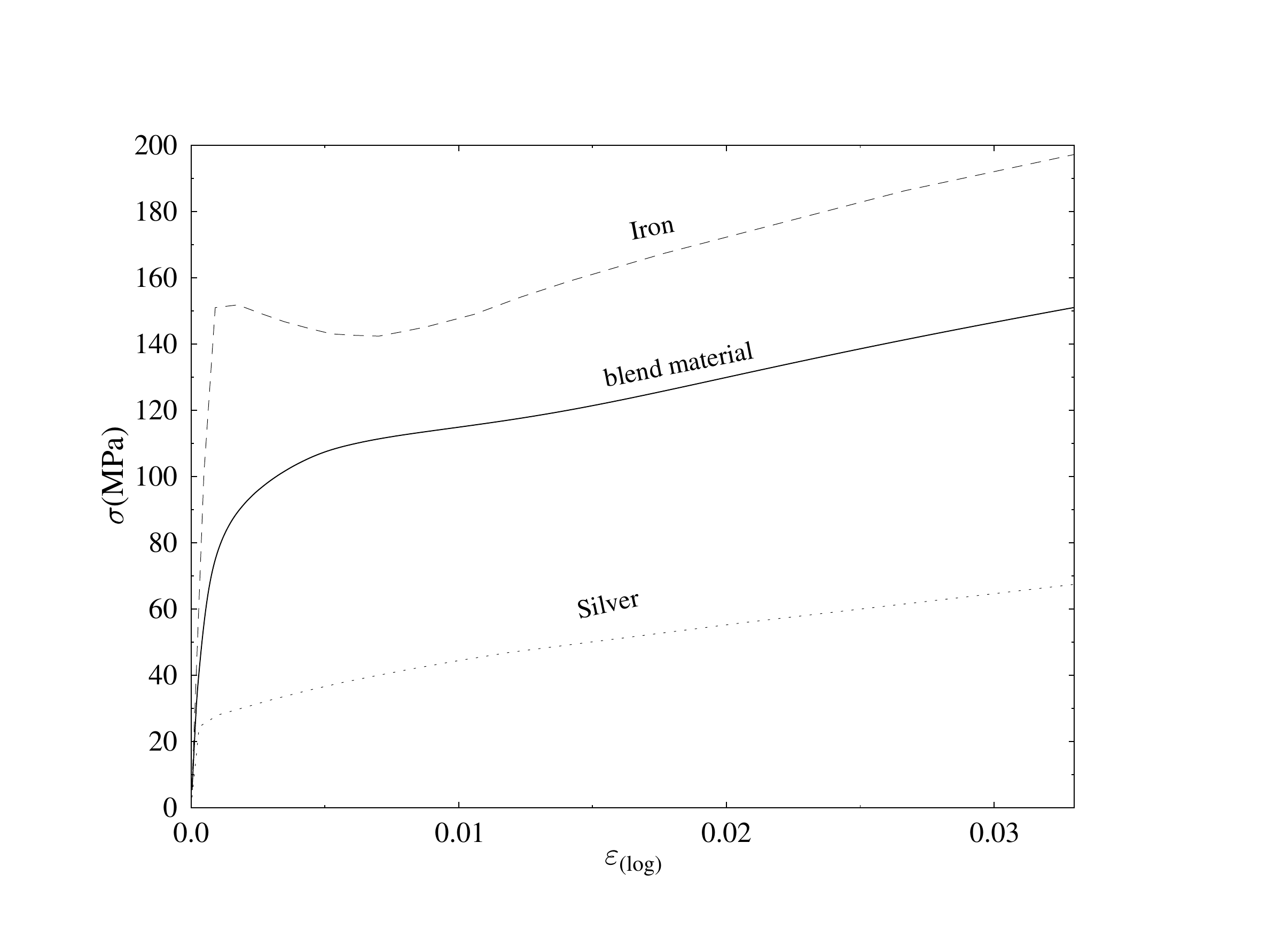}
      \vskip 0.5cm
      \begin{center}
        {\it (c)} 
      \end{center}
    }
  }
  \caption{{
      (a): Microstructure of a silver/iron blend material observed by 
      Scanning Electron Microscopy. (b): response of the individual constituents
      under uniaxial tension.
      (c): numerical simulation. Uniaxial tension in the horizontal 
      direction. Overall strain $E_{11}= 3.3 \%$. Maps of equivalent plastic strain.
  }}
  \label{bornert}
\end{figure}


\clearpage
%
\begin{table}[!p]
  \begin{center}
    \begin{tabular}{|c|cc|cc|}
      \hline
      & \multicolumn{2}{c|}{\rm Square arrangement}
      & \multicolumn{2}{c|}{\rm Hexagonal arrangement} \\
      Resolution & Young's \ modulus & Error \ (\%)
      & Young's \ modulus & Error \ (\%) \\
      \hline
      & & & & \\
      32   & 153 \ 965.  &     0.385 &   140 \ 040. &    0.28  \\ 
      64   & 153 \ 055.  &    -0.209 &   140 \ 130. &    0.34  \\
      128 & 153 \ 385.  &     0.007 &   139 \ 535. &    -0.08  \\
      256 & 153 \ 150.  &    -0.147 &   139 \ 580. &    -0.05  \\
      512 & 153 \ 145.  &    -0.150 &   139 \ 655. &    0.00  \\
      1024  & 153 \ 190.  &    -0.121 &   *          &    *   \\
      2048  & 153 \ 375.  &     0.000 &   *          &    *    \\
      & & & & \\
      \hline
    \end{tabular}
  \end{center}
  \caption{Square and hexagonal array. Transverse tension 
    at $0^0$. Influence of spatial resolution on 
    the overall Young's modulus. 
  }
  \label{tab1} 
\end{table}


\begin{table}[!p]
  \begin{center}
    \begin{tabular}{|c|cc|cc|}
      \hline
      & \multicolumn{2}{c|}{\rm Square arrangement}
      & \multicolumn{2}{c|}{\rm Hexagonal arrangement} \\
      Resolution & Young's \ modulus & Error \ (\%)
      & Young's \ modulus & Error \ (\%) \\
      \hline
      & & & & \\
      32   & 129 \ 670.  &    0.83   &   140 \ 810. &    0.88  \\
      64   & 128 \ 400.  &   -0.16   &   140 \ 200. &    0.44  \\
      128  & 128 \ 750.  &    0.12   &   139 \ 680. &    0.07  \\
      256  & 128 \ 660.  &    0.05   &   139 \ 520. &   -0.04  \\
      512  & 128 \ 600.  &    0.00   &   139 \ 580. &    0.00  \\
      & & & & \\
      \hline
    \end{tabular}
  \end{center}
  \caption{
    Square and hexagonal array.
    Transverse tension at $45^\circ$.
    Influence of spatial resolution on the overall Young's modulus.}
  \label{tab2}
\end{table}


\begin{table}[!p]
  \begin{center}
    \begin{tabular}{|c|cc|cc|}
      \hline
      & \multicolumn{2}{c|}{\rm Square arrangement}
      & \multicolumn{2}{c|}{\rm Hexagonal arrangement} \\
      Resolution & Flow\ stress & Error \ (\%)
      & Flow\ stress & Error \ (\%) \\
      \hline
      & & & & \\
      32  & 112.39  & 15.04  & 88.48   & 0.60 \\
      64  & 107.46  & 9.99   & 88.32   & 0.42 \\
      128  & 102.29  & 4.70   & 88.10   & 0.18 \\
      256  &  99.65  & 2.00   & 88.01   & 0.07 \\
      512  &  98.61  & 0.93   & 87.95   & 0.00 \\
      1024 &  98.01  & 0.32   & *       & *    \\
      2048 &  97.70  & 0.00   & *       & *    \\
      & & & & \\
      \hline
    \end{tabular}
  \end{center}
  \caption {
    Square and hexagonal array. 
    Transverse tension at $0^\circ$.
    Influence of spatial resolution on the overall flow stress.}
  \label{tab3}
\end{table}

\clearpage

\begin{table}[!p]
  \begin{center}
    \begin{tabular}{|c|cc|cc|}
      \hline
      & \multicolumn{2}{c|}{\rm Square arrangement}
      & \multicolumn{2}{c|}{\rm Hexagonal arrangement} \\
      Resolution & Flow\ stress & Error \ (\%)
      & Flow\ stress & Error \ (\%) \\
      \hline
      & & & & \\
      32  &  79.558  & 0.00   & 79.554  & 0.00 \\
      64  &  79.558  & 0.00   & 79.554  & 0.00 \\
      128  &  79.558  & 0.00   & 79.554  & 0.00 \\
      256  &  79.558  & 0.00   & 79.554  & 0.00 \\
      & & & & \\
      \hline
    \end{tabular}
  \end{center}
  \caption{
    Square and hexagonal array. 
    Transverse tension at $45^\circ$.
    Influence of spatial resolution on the overall flow stress.
  }
  \label{tab4}
\end{table}


\begin{table}[!p]
  \begin{center}
    \begin{tabular}{|c|cc|cc|}
      \hline
      & \multicolumn{2}{c|}{\rm Square arrangement}
      & \multicolumn{2}{c|}{\rm Hexagonal arrangement} \\
      Resolution & Hardening modulus & Error \ (\%)
      & Hardening modulus & Error \ (\%) \\
      \hline
      & & & & \\
      
      32  &  14.4 \ $10^3$ &  7.46  & 7.50 \ $10^3$  & 5.63 \\
      64  &  13.8 \ $10^3$ &  2.99  & 7.30 \ $10^3$  & 2.82 \\
      128  &  13.6 \ $10^3$ &  1.49  & 7.10 \ $10^3$  & 0.00 \\
      256  &  13.4 \ $10^3$ &  0.00  & 7.10 \ $10^3$  & 0.00 \\
      512  &  13.4 \ $10^3$ &  0.00  & 7.10 \ $10^3$  & 0.00 \\
      & & & & \\
      \hline
    \end{tabular}
  \end{center}
  \caption{
    Square and hexagonal array. Transverse tension at $0^\circ$.
    Influence of spatial resolution on the overall hardening modulus.}
  \label{tab5}
\end{table}


\begin{table}[!p]
  \begin{center}
    \begin{tabular}{|c|cc|cc|}
      \hline
      & \multicolumn{2}{c|}{\rm Square arrangement}
      & \multicolumn{2}{c|}{\rm Hexagonal arrangement} \\
      Resolution & Hardening modulus & Error \ (\%)
      & Hardening modulus & Error \ (\%) \\
      \hline
      & & & & \\
      32  &  4.94 \ $10^3$ &  3.72  & 7.94 \ $10^3$  & 7.01 \\
      64  &  4.78 \ $10^3$ &  0.42  & 7.62 \ $10^3$  & 2.70 \\
      128  &  4.78 \ $10^3$ &  0.42  & 7.50 \ $10^3$  & 1.08 \\
      256  &  4.78 \ $10^3$ &  0.42  & 7.44 \ $10^3$  & 0.27 \\
      512  &  4.76 \ $10^3$ &  0.00  & 7.42 \ $10^3$  & 0.00 \\
      & & & & \\
      \hline
    \end{tabular}
  \end{center}
  \caption{
    Square and hexagonal array. Transverse tension at $45^\circ$.
    Influence of spatial resolution on the overall hardening modulus.}
  \label{tab6}
\end{table}

\clearpage

\begin{table}[!p]
  \begin{center}
    \begin{tabular}{|c|c|c|c|c|}
      \hline
      Number \ of & Number \ of & Young's \ modulus
      & standard & error on   \\
      fibers & tests &  mean (GPa) &   deviation (GPa) 
      & mean \ ($\%$)\\
      \hline
      & & & & \\
      4   & 100 & 143.7  & 3.9    & 0.27 \\
      9   & 50  & 143.4  & 3.1    & 0.30 \\
      16   & 40  & 143.0  & 2.6    & 0.29 \\
      36   & 25  & 143.1  & 1.51   & 0.21 \\
      64   & 27  & 143.2  & 1.33   & 0.19 \\
      256  & 10  & 142.9  & 0.57   & 0.13 \\
      & & & & \\
      \hline
    \end{tabular}
  \end{center}
  \caption{
    Random configurations. Transverse uniaxial tension 
    in the horizontal direction.
    Influence of the size of the unit cell on the overall Young's modulus.
  }
  \label{tab7}
\end{table}


\begin{table}[!p]
  \begin{center}
    \begin{tabular}{|c|c|c|c|c|}
      \hline
      Number \ of & Number \ of & Flow \ stress
      & standard & Error  \\
      fibers & tests &  mean (MPa) &  deviation (MPa) & on \ mean (\%)\\
      \hline
      & & & & \\
      4   & 100 &    89.54  &   6.07 & 0.68 \\
      9   & 50  &    88.01  &   5.04 & 0.81 \\
      16   & 40  &    87.94  &   4.99 & 0.90 \\
      36   & 25  &    88.15  &   2.17 & 0.49 \\
      64   & 27  &    88.70  &   2.07 & 0.51 \\
      256  & 10  &    88.88  &   0.64 & 0.23 \\
      & & & & \\
      \hline
    \end{tabular}
  \end{center}
  \caption{
    Random configurations. Transverse uniaxial tension 
    in the horizontal direction.
    Influence of the size of the unit cell on the overall flow stress.
  }
  \label{tab8}
\end{table}


\begin{table}[!p]
  \begin{center}
    \begin{tabular}{|c|ccc|}
      \hline
      Fiber shape &Young's Modulus & Flow stress & Hardening modulus \\
      & mean (MPa)          & mean (MPa)       & mean (Mpa) \\
      \hline
      & & & \\
      circle   & 142 \ 260 &   86.9 &   9 \ 382  \\
      triangle & 142 \ 250 &   91.4 &  10 \ 448  \\
      ellipse  & 142 \ 330 &   88.7 &   9 \ 180  \\
      & & & \\
      \hline
    \end{tabular}
  \end{center}
  \caption{
    Random configurations. Transverse uniaxial tension 
    in the horizontal direction.
    Effect of the shape of fibers on the effective properties
    of the composite.}
  \label{tab9}
\end{table}

\clearpage

\begin{table}
  \begin{center}
    \begin{tabular}{|c|ccc|}
      \hline
      Fiber shape & Maximal length & Minimal length & Average length  \\
      \hline
      & & & \\
      circle   & 1    &  1    & 1     \\
      triangle & 1.35 & 1.17  & 1.29  \\
      ellipse  & 1.72 & 0.52  & 0.97  \\
      & & & \\
      \hline
    \end{tabular}
  \end{center}
  \caption{
    Projections along different angles of fibers with
    the same surface $s=1$.}
  \label{tab10}
\end{table}

\newpage
\bibliographystyle{unsrt}
\bibliography{books,articles}
\end{document}